\documentclass[runningheads]{llncs}
\usepackage[T1]{fontenc}
\usepackage{graphicx}
\usepackage{xcolor}
\definecolor{DSgray}{cmyk}{0,1,0,0}

\usepackage{subfigure}

\usepackage{xurl}
\usepackage{cite}
\usepackage{multirow}
\usepackage[figuresleft]{rotating}
\usepackage{hyperref}

  \makeatletter
    \def\@fnsymbol#1{\ensuremath{\ifcase#1\or  \dagger\or \ddagger
    \or \mathsection \or * \or \mathparagraph\or \|\or **\or \dagger\dagger
   \or \ddagger\ddagger \else\@ctrerr\fi}}
    \makeatother
    
\begin{document}
\title{SoK: Play-to-Earn Projects
}
\author{Jingfan Yu\inst{1}\thanks{\email{yujf20@mails.tsinghua.edu.cn}}
\and
Mengqian Zhang\inst{2}\thanks{\email{mengqian.cs@gmail.com}}
\and
Xi Chen\inst{3}\thanks{\email{xchen3@stern.nyu.edu}}
\and
Zhixuan Fang\inst{1,4}\thanks{Corresponding author: Zhixuan Fang 
(\email{zfang@mail.tsinghua.edu.cn})}
}
\authorrunning{J. Yu et al.}
\institute{Tsinghua University, Beijing, China\\
\and
Shanghai Jiao Tong University, Shanghai, China\\
 \and
New York University, New York, USA\\
\and
Shanghai Qi Zhi Institute, Shanghai, China\\
}

\maketitle              % typeset the header of the contribution

\begin{abstract}
% The abstract should briefly summarize the contents of the paper in
Play-to-earn is one of the prospective categories of decentralized applications. The play-to-earn projects combine blockchain technology with entertaining games and finance, attracting various participants. While huge amounts of capital have been poured into these projects, the new crypto niche is considered controversial, and the traditional gaming industry is hesitant to embrace blockchain technology. In addition, there is little systematic research on these projects. In this paper, we delineate play-to-earn projects in terms of economic \& governance models and implementation and analyze how blockchain technology can benefit these projects by providing system robustness, transparency, composability, and decentralized governance. We begin by identifying the participants and characterizing the tokens, which are products of composability. We then summarize the roadmap and governance model to exposit there is a transition from centralized governance to decentralized governance. We also classify the implementation of the play-to-earn projects with different extents of robustness and transparency. Finally, we discuss the security \& societal challenges for future research in terms of possible attacks, the economics of tokens, and governance.

\keywords{Play-to-Earn \and Gamefi \and Tokenomics.}
\end{abstract}
\section{Introduction}
The development of blockchain, especially the emergence of fungible tokens(FTs) and non-fungible tokens (NFTs), is seeing big changes in the business model of gaming industries. By giving value back to the players, the novel play-to-earn (P2E) model has attracted huge attention from the market. In September 2022, there are over 900 thousand active wallets interacting with blockchain-based gaming projects and over 20 million transactions are sent~\cite{dappradar}.

The earliest blockchain-based gaming projects date back to 2015. These projects store the ownership of players' in-game assets in blockchains to provide crypto-assets to the players. Launched in 2015, Spells of Genesis~\cite{spellsofgenesis} is the 1st blockchain-based mobile game and is partnered with Counterparty~\cite{counterparty}. In the game, players have to collect cards and fight their enemies. Players interact with the game server governed by the creator EverdreamSoft, except that there is a `blockchainization' button that converts in-game cards to blockchain-based digital assets. When a player with digital wallets of the corresponding blockchain chooses a card and clicks the button, the server will send a transaction to the blockchain to record the ownership of the card on it.
Launched in 2017, CryptoKitties is a successful game based on Ethereum smart contract~\cite{wood2014ethereum}, hitting over 14,914 users a day at its peak~\cite{CryptoKittiesPeak}. In the game, players can trade and breed their CryptoKitties through the smart contract by sending transactions to the Ethereum network. 
% Utilizing the custumizable feature of smart contracts~\cite{wang2018overview}, the smart contracts of the project not only record the ownership of the CryptoKitties, but also record genetic mechanisms and trace the breeding process.
% Smart contracts are codes executed on blockchains, and those running on Ethereum are written in Turing-complete programming language so that the creator can customize the contract with various functions other than transferring assets ~\cite{wang2018overview}.

The development of token standards and decentralized finance (Defi)~\cite{werner2021sok} lead to the emergence of various token designs attracting various investors, \textit{e.g.} MakerDAO~\cite{makerdao} established its stablecoin~\cite{moin2020sok} in 2017. Defi projects affect the designs of blockchain-based games and also provide secondary markets for game assets trading. The combination of Defi and game is also called Gamefi.
Launched in 2018, Axie Infinity~\cite{axie} is one of the most popular Gamefis, hitting a total NFT transaction volume of over \$2 billion by September 2021~\cite{AxieStorm}. It is a battling game and allows players to collect, raise, breed, battle, and trade digital pets known as Axies. The game attracted 2.8 million monthly active users at its peak by rewarding players with tokens~\cite{AxiePeak}. 
% Axies are tokenized as NFTs. There are two main FTs used in the Axie Infinity universe: Axie Infinity Shards (AXS) and Smooth Love Potions (SLP). A mini-documentary called 'play-to-earn'\footnote{https://www.youtube.com/watch?v=Yo-BrASMHU4} interviewed players of the game, making the word 'play-to-earn' popular.
% Some of the players in Philipin earn their livings by playing the game. Many Gamefi projects emphasize their 'play-to-earn' fe
% \footnote{https://www.bloomberg.com/news/articles/2021-12-07/play-to-earn-crypto-gaming-apps-surpass-defi-in-user-popularity}
% (or AXS, which is the ICO token)(or SLP, which is the utility token)
On the other hand, the high return and the high entry cost incur criticism that there are economic bubbles in these projects, and some may consider these projects as Crypto Ponzi schemes. Therefore, many projects also offer a free-to-play mechanism to reduce the entry cost, \textit{e.g.}, Illuvium and a new version of Axie called Axie Origin~\cite{axie}. In addition, the transparency of the blockchain may mitigate financial fraud.
% Ponzi scheme contains cheating. T

A P2E project is a typical Gamefi project where players can earn crypto-assets depending on the engagement level in the game as its rule suggests. The rule is called game \textit{protocol} in this paper and is created by the project founder, \textit{e.g.}, in CryptoKitties, the entrepreneur creates the smart contract. The in-game assets like virtual land, characters, as well as skins, are in the form of crypto tokens, either FTs or NFTs. Also, there can be tokens for the monetary system of the game, such as gems, points, or coins, with which players can purchase the assets. The information of ownership, transaction activities of these tokens are stored on blockchains, which could be public blockchains like Ethereum, or program-specific chains with preferred features. 
% Besides the projects mentioned above,
Other than entertaining games mentioned above, some projects called Metaverse~\cite{vidal2022new}, X-to-earn\footnote{There are move-to-earn and read-to-earn projects, such as stepN~\cite{stepn} and ReadON~\cite{readon}.} or crypto-game ecosystems are also included in our definition of P2E projects, as long as there is a rule or a protocol and players can earn crypto-assets by taking actions defined in the protocol.
% 'Metaverse' and 'X-to-earn', including move-to-earn and read-to-earn, For example, StepN~\cite{stepn} is called move-to-earn because players have to run in the real world when playing the game. Game platform that contain multiple games They can also be considered as one game 

The P2E project is different from the traditional game in terms of the following properties:
\begin{itemize}
    \item \textbf{Robustness against single-point failure:} any single party can not affect the state of the game. For example, malicious behaviors of the entrepreneur or the failure of a single entity's servers that execute the game will not change the asset ownership of players.
    \item \textbf{Transparency:} the execution of the smart contract is tractable by anyone.
    \item \textbf{Composability}~\cite{werner2021sok} to secondary markets: the function of the smart contract of tokens is callable by other contracts so that any standard token can be traded in secondary markets.
    \item \textbf{Decentralized governance:} the updates of the game are decided by the players instead of the project founder.
\end{itemize}
%  Robustness against single-point failure means  Transparency means  Composability means that Decentralized governance means 
 
Traditional games are played in a centralized ecosystem where developers retain almost all power over the experience. Players devote their time and effort to their favorite games and collect various characters, weapons, extra powers, and more. These achievements, however, only value within games. When the game ends, the precious collections will be worthless. People cannot transfer their hard-earned items to another game, exchange them on third-party marketplaces or simply cash them out. In a traditional game model, players do not own the value they created. %In traditional games, players are subject to the decisions of developers and game companies.
% level of engagement
% trade in blockchain
% entrepreneur nodes
By contrast, in a P2E project, the four features mentioned above make the tokens valuable outside of the game and realize true ownership of tokens for the players.
These features depend on the implementation and token designs, which will be discussed in the remaining of the paper.
% robustness, no matter the behaviors of the game's developers, and can trade them for real money or other cryptocurrencies at any time.
% In addition, players can participate in the update of the game. They can put forward proposals for the update and vote for proposals. Their voting is usually weighted by the aforementioned tokens they possess. 

% marketcap 
% history
% elements definition
% challenge

% NFT  In Ethereum, ERC-20 is the FT standard and .
To the best of our knowledge, this paper is the first systematical paper to review P2E projects. We identify the participants, characterize the economic \& governance model, and summarize various purposes of tokens. We discuss the technical roadmap and implementation realization and propose a classification of popular projects. We discuss the security \& societal challenges of the projects. The rest of the paper is organized as follows: In Section~\ref{sec:Play-to-Earn Games}, we discuss the economic \& governance model. In Section~\ref{sec:Implementation}, we discuss the implementation, especially how the data are stored and processed, which reflects different design principles. In Section~\ref{sec:Challenges}, we discuss the challenges and solutions.

\section{Play-to-Earn Games}
\label{sec:Play-to-Earn Games}
In this section, we discuss the economic \& governance model. In Section~\ref{subsec:Participants}, we identify the participants. In Section~\ref{subsec:Tokens}, we summarize the tokens in terms of their purposes. Section~\ref{subsec:Tokenomics} summarizes the flow of tokens in P2E projects and Section~\ref{subsec:Roadmap} discusses their development roadmap. In Section~\ref{subsec:Decentralized Governance}, we discuss the decentralized governance of P2E projects.

\subsection{Participants}
\label{subsec:Participants}
In this paper, we call the world in the game \textit{universe} and focus on three mainly involved parties of a P2E game: entrepreneur, investor, and player.

\begin{itemize}
    \item The \textit{entrepreneur} is an individual or a team who designs the game protocol and launches the game universe. By realizing a series of new ideas, the entrepreneur aggregates capital and provides the game service for profit.
    \item An \textit{investor} is the person who provides the game project with capital in the form of fiat money or cryptocurrency like Bitcoin. Meanwhile, the investor receives the token issued by the game. This investment process is mainly done through initial coin offerings (ICOs), which is a popular way for the entrepreneur to raise funds in the blockchain. Usually, investors do so out of the expectation that the received tokens will appreciate and bring high returns for them in the future. % They can also earn rewards by staking these tokens in the game. 
    \item A \textit{player} is the user of the game. He engages in activities allowed in the game ecosystem for entertainment. In this process, the player can obtain various tokens in the game. 
\end{itemize}
The players and the investors are differentiated by whether they play the game, or say, take actions that are defined by the game protocol. Investors will buy tokens before the game protocol is created, while the players won't.

\subsection{Tokens}
\label{subsec:Tokens}
In this subsection, we discuss different token designs in P2E projects. For example, some projects use the same fungible token for funding and incentivizing players, while others like Axie Infinity~\cite{axie} and stepN~\cite{stepn} use dual fungible tokens.

There are usually many kinds of tokens in a P2E game. The aforementioned token involved in ICOs (also referred to as \textit{ICO token} in this paper) is fungible and its total circulation is limited or scheduled, \textit{e.g.}, AXS of Axie Infinity~\cite{axie}. Besides, the game may issue other fungible tokens, \textit{e.g.}, SLP of Axie Infinity (which is called \textit{utility token}), mainly for players' use in the game. Its total supply can be unlimited. In addition, some NFTs will be generated during the play. All tokens in a game including the ICO token, utility token, and NFTs are collectively referred to as \textit{project tokens}. The possible design purposes of project tokens are summarized as follows.
\begin{itemize}
    \item \textbf{Funding:} to raise money for creating or updating the game universe. For example, ICO tokens can work as funding tokens.
    \item \textbf{Incentive for players:} to reward players for participating in games (\textit{e.g.}, winning a tournament, or creating content).
    % specified behaviors
    \item \textbf{Payment:} for players to pay to the universe for assets or playing games.
    \item \textbf{Voting:} for voters to participate in the voting when governing the universe. Tokens with the voting function are also called \textit{governance token}.
    % The tokens used to calculate the voters' weight when voting. The tokens are also called \textit{governance token}.
    \item \textbf{Incentive for investors:} to reward investors for their staking of funding tokens they bought.
    \item \textbf{Tokenizing user-generated content (UGC):} to represent user-generated content and make the content tradable as assets. It is usually the non-fungible tokens that achieve this function.
    \item \textbf{Incentive for implementation:} to reward users for their implementation of the project. For example, for the project using project-specific consensus, miners receive the tokens with this purpose as the reward for mining.
    % that relies on users to realize its implementation
    \item \textbf{Stake proof for consensus:} to elect leaders or voters when adopting Proof-of-Stake (PoS)~\cite{saleh2021blockchain,kiayias2017ouroboros} for the consensus.
\end{itemize}
Table~\ref{tab:token} lists the design of token purposes in several popular P2E projects. As can be seen, each project can be characterized by its token designs, \textit{e.g.}, whether to use different tokens as the incentive for players and investors, whether the governance token can be earned by playing, and whether there are tokens that cannot be earned by playing. In addition, we can see that one token can be used for multi-purpose~\cite{oliveira2018token}, and there can be multiple tokens for the same purpose. This allows game designers to combine tokens with various features to design distinctive projects and deal with complex game ecology. 

\subsection{Tokenomics}
\label{subsec:Tokenomics}
In this subsection, we summarize three mechanisms that exist in most P2E projects and introduce the overall flow of tokens.

\noindent\textbf{Minting Mechanism.}
The minting mechanism is a set of rules allowing players to generate new tokens in the game. This can require players' efforts and consume players' tokens. For example, in Axie Infinity, a player consumes its AXS and SLP to breed new Axies NFTs. After mining an NFT, players can do more other than just own it in the universe. They can use the NFT to participate in various events to earn more project tokens, rent them to other players for revenue, or sell them in the market.

\noindent\textbf{Staking Mechanism.}
Staking is a way for investors to earn in the game. Although the players can earn this way, we mainly consider the earnings of the investors because if players stake all their tokens, they cannot play the game. A staking mechanism requires token holders to lock their project tokens for a while and promises to pay them interest for this. The interests are also in the form of project tokens and proportional to the number of locked tokens. 
% For the game itself, the development of the ecosystem can be adjusted in this way. Specifically, in the early ICO phase, 
This could serve as a bright spot to attract investors who 
%allocate their capital with the expectation of getting a profit in the future. Although investors typically don't invest the time and effort to play the game and thus can not enjoy the benefits of playing to earn, they 
can gain wealth by staking to earn. In addition, % staking is a powerful tool to incentivize investors. 
regarding staking as an approach to money-making money, it can also prevent token holders from overselling their tokens by increasing interest rates. 
% Generally speaking, if the return on staking is increased, more tokens will be locked, which will decrease the token in circulation, and make the tokens more valuable. However, the staking interest is not created out of thin air. 
Typically, the system reserves a certain amount of tokens in advance as staking interest (stated in the whitepaper), as well as collecting assets from transaction fees and token flows in the universe (\textit{e.g.}, consumed tokens for minting NFTs).

\noindent\textbf{Treasury Mechanism.}
To develop a sustainable economy, the treasury mechanism is designed to collect assets from the universe and then allocate them to certain entities. The treasury funds are primarily in the form of fees from game-related economic activities. Specifically, when consuming tokens to mint NFTs in the game, part of them will be recycled into the treasury. Besides, when a successful NFT sale occurs in the marketplace, some transaction fees will be extracted from the trade. Also, in some activities where the entrepreneur makes money (\textit{e.g.}, advertisement and subscription service), part of the revenue will be put into the treasury. Later, treasury funds are allocated to the ecosystem according to some rules. For example, they can serve as the staking interest or monetary rewards to encourage active contributions to the game's growth.

\begin{figure}
    \centering
    \includegraphics[width=0.6\linewidth]{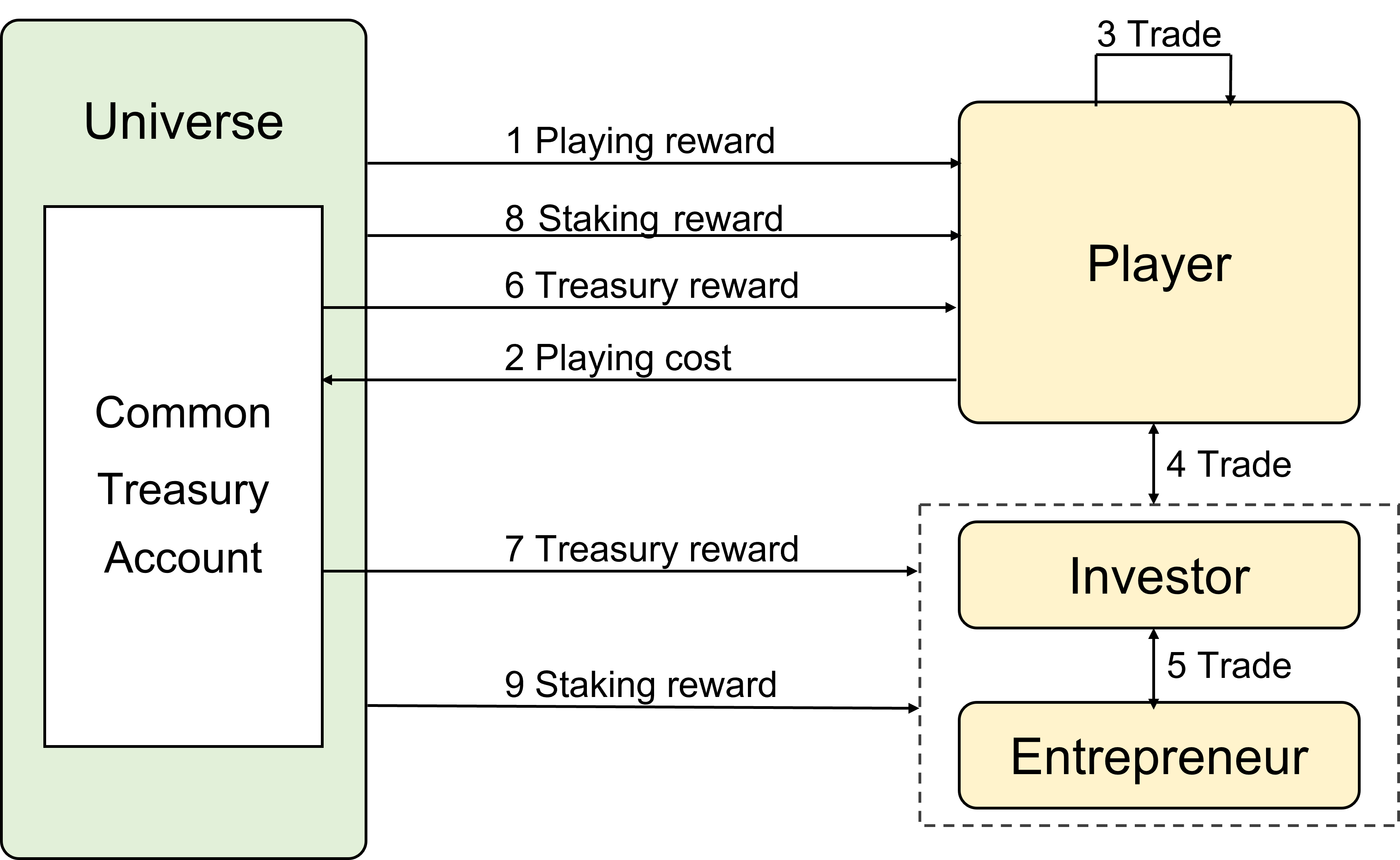}
    \caption{The flow of tokens.}
    \label{fig:token_flow}
\end{figure}

\noindent\textbf{Flow of Tokens.}
In this paragraph, we describe the flow of tokens as well as currency in a P2E game, which is illustrated in Figure~\ref{fig:token_flow}. First, as Line 1 shows, players can gain project tokens or currency through participating in the universe, \textit{i.e.}, taking actions specified by the universe. As a cost, this process also consumes some tokens (see Line 2). Those project tokens paid to the universe will either go to a treasury account for further use or be transferred to a burn address where the token can never be retrieved. Moreover, it may be necessary to hold an NFT when playing. For example, in Axie Infinity, players can earn AXS and SLP by participating in tournaments and attaining top rankings. To join the competition, players should have an Axie NFT in advance. For a new player without such an NFT, he can choose to buy one from other players, using project tokens, fiat money, or cryptocurrencies like ETH. Actually, all project tokens can be traded by players, investors, and the entrepreneur in the marketplace (see Lines 3, 4, and 5). % For example, players can buy some ICO tokens from the holders such as investors. 
As mentioned, part of the tokens or currency involved in the trade may go to the treasury in the form of transaction fees. In Axie Infinity, ETH can be used for trading an Axie NFT, and after a trade, 4.25\% ETH will be paid to the Axie Common Treasury. Those collected project tokens and currencies in the treasury can be distributed to some token holders (see Lines 6 and 7). The distributed method is originally specified in the whitepaper or later decided by the token owners through weighted voting. For example, when Axie Infinity evolves into the period of decentralized organization, AXS holders will vote for the rules on how to distribute AXS and ETH in the Axie Common Treasury to the community.

% In addition, part of the tokens in this common account and the staking funds will be allocated to 
Lines 8 and 9 show the staking process that distributed tokens to particular token holders. In this process, this particular token needs to be staked in a dashboard for some specified period to get the reward, and can not be used for trading during the period. The token portfolio earned can be different depending on what token is staked and how long the token is staked.
\subsection{Roadmap}
\label{subsec:Roadmap}
\begin{figure*}[!t]
    \centering
    \includegraphics[width=\textwidth]{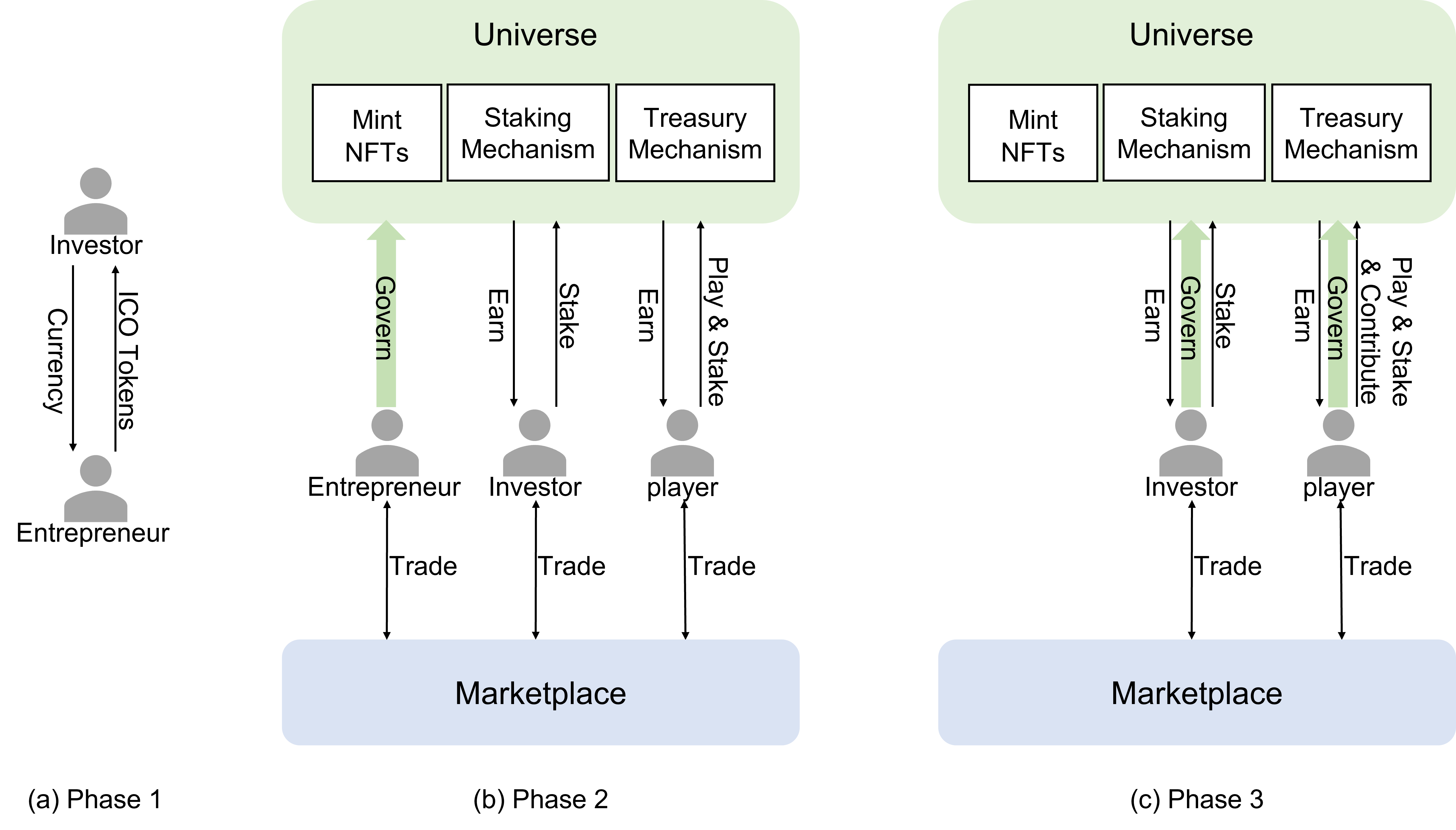}
    \caption{Roadmap of P2E projects.}    
    \label{fig:1}   
\end{figure*}
In this subsection, we present the evolution of P2E games, which are built from scratch and then change from centralization to decentralization with the flow of tokens. Most P2E games plan to delegate the governing power to the community, using the project token as a medium. Holders of the governance token will have the right to participate in the decision-making of a certain event. Besides the ICO token, NFTs or other project tokens may be also regarded as the governance token, which varies among different P2E games. 

At the very beginning, the entrepreneur is the only token holder. During the ICO stage, some ICO tokens are sold to the investors for funds, as shown in Figure~\ref{fig:1}(a). Along with the token itself, these investors also have the potential rights to manage this game project in the future, if the ICO token is regarded as the governance token. Then in Phase 2, tokens are further dispersed among participants. Players also own the governance token. But in this phase, the game universe is still mainly managed by the entrepreneur. Ideally, the system will eventually evolve to Phase 3, namely, it is completely managed by the governance token holders including players and investors. They can vote for the improvement of the game. For example, they may vote on whether to pass a proposal about adding a new feature to the game. If the proposal passed, they also need to decide the amount of reward for attracting someone to implement this function. Thus, in Phase 3, the players can also earn money by contributing to the universe. In addition, the original entrepreneur acts as a player or investor in this phase. Likewise, it can obtain tokens by playing, contributing, or just staking, and it can also participate in governance.

\subsection{Decentralized Governance}
\label{subsec:Decentralized Governance}
% DAO is essential if the project claims that players can take control of their assets in the universe. 
Most P2E projects will gradually evolve into the phase that uses Decentralized Autonomous Organization (DAO) to govern the universe so that players can make decisions for the development of the universe. They can vote for specified parameters or proposals proposed by the community. Their votes are usually weighted by their governance tokens either linearly or quadratically. Various rules, such as the votes threshold to pass a proposal, starting and ending time of voting, and voting rewards are initially decided by the entrepreneur and then updated by the DAO. If multiple tokens are allowed for voting, their relative weight should be specified. In addition, participants' behaviors may also influence their voting weight. For example, it can be weighted by the staking period.

% Votes can also be weighted directly by behaviors. For example, it can be weighted by the staking period.

The governance tokens can be gained through various behaviors in the universe, reflecting different design principles of the project. If a project prefers the longtime or skillful players to be governors, it can make the tokens mainly maintained by higher-level players have a larger weight when voting. For example, GMT, which is the governance token in Stepn, can only be gained when reaching level 50 and Axie Infinity's governance token AXS is distributed to high-rank players in tournaments.

\section{Implementation}
\label{sec:Implementation}
In this section, we discuss various implementations of the P2E projects. We focus on how the state of the game universe is stored and updated upon receiving the actions of the participants, where states are the view of the game universe shared by all the participants. 
% are calculated from all the input data by Turing machines.
Different implementations reflect the design principles of the project, \textit{e.g.}, decentralization, transparency, security, and performance. In Tabel~\ref{tab:implementation_brief}, we briefly list the implementations of the projects mentioned in this section and summarize the implementations of several popular P2E projects in Table~\ref{tab:implementation} in more detail. We discuss the principles behind these designs in this section.

% In Table~\ref{tab:implementation}, we list the implementations of the projects. We will discuss how these different implementations reflect different principles in this section.

The structure of this section is as follows.
First, we discuss decentralized data storage in Section~\ref{subsec:Decentralized data storage}. The data in P2E projects can be stored in either a central server or decentralized nodes connected by P2P networks. Data stored in central servers are vulnerable to single-point failure, \textit{i.e.}, the entity fails or becomes malicious. In addition, data operations are not transparent. On the other hand, data stored in decentralized nodes may suffer from inconsistency and poor performance. Therefore, consensus protocols are used to resolve conflict states, which is discussed in Section~\ref{subsec:Consensus}. In Section~\ref{subsec:Bridges and oracles}, we discuss bridges and oracles, which are used for communication between separate components.
% To improve the performance, data are stored differently depending on whether they are necessary for the consensus of the future state. The methods of decentralized data storage are discussed in Section 2.3.
% 
% we discuss methods to solve these problems by dealing the following sections. 
% consensus: how
% decentralized storage: big raw data that is unnecessary for future state
In Section~\ref{subsec:Hybrid architecture}, we classify these implementations into fully decentralized and hybrid architectures, and discuss the design principles of different implementations.
% in terms of the management and storage realization of data and states
% \begin{figure}
%     \centering
%     \includegraphics[width=\linewidth]{figures/implementation_table.png}
%     \caption{The implementations of the P2E projects that are mentioned in this section.}
%     \label{fig:implementation_table}
% \end{figure}
\begin{table}[h!]
\caption{The implementations of P2E projects mentioned in Section~\ref{sec:Implementation}.} \label{tab:implementation_brief}
\begin{minipage}{\textwidth}      
\scriptsize
  \begin{center}
    \begin{tabular}{|c|c|c|c|c|c|c|c|}
        \hline
        \multirow{2}{*}{projects} & \multirow{2}{*}{architecture} & \multirow{2}{*}{data} & \multirow{2}{*}{implementation} & public& permission & consensus  & \multirow{2}{*}{bridge\footnote{bridge: Y means the project designs its own bridge.}}\\
         &  & & &(Y/N)\footnote{public: Y means public, N means program-specific.}& (Y/N)\footnote{permission: Y means permissioned, N means permissionless.} & protocol &\\
        \hline
        My Neighbor& \multirow{2}{*}{decentralized} & \multirow{2}{*}{all\footnote{including asset ownership (AO), UGC, game protocol \& its execution (GP).}} & \multirow{2}{*}{Chromia~\cite{chromia}} & \multirow{2}{*}{Y} & \multirow{2}{*}{Y} & \multirow{2}{*}{BFT} & \multirow{2}{*}{-}\\
        Alice~\cite{myneighboralice} &  &  &  &  &  &  & \\
        %  &  &  & anchor blocks in PoW chains &  & \\
        \hline
        \multirow{2}{*}{Gala Games~\cite{gala}} & \multirow{2}{*}{hybrid} & AO & Ethereum~\cite{wood2014ethereum} & Y & N & Nakamoto & \multirow{2}{*}{-}\\
         & & others\footnote{other game content.} & Gala Node~\cite{galanode} & N & N & monitored\footnote{monitored means monitored by Gala Game.}& \\
        %   &  & & &  &  & (entrepreneur)&\\
        \hline
        \multirow{3}{*}{The Sandbox~\cite{sandbox}} & \multirow{3}{*}{hybrid} & AO & Ethereum~\cite{wood2014ethereum}
        & Y & N & Nakamoto & -\\
         &  & UGC & IPFS~\cite{benet2014ipfs} & - & - & -&-\\
          &  & GP & AWS & - & - & -&-\\
        \hline
        \multirow{3}{*}{Decentraland~\cite{decentraland}} & \multirow{3}{*}{hybrid} & AO & Ethereum~\cite{wood2014ethereum}
        & Y & N & Nakamoto & -\\
         &  & UGC & BitTorrent/IPFS & - & -& - & -\\
         &  & GP & User front-end & - & -& - & -\\
         \hline
        \multirow{2}{*}{Axie Infinity~\cite{axie}} & \multirow{2}{*}{hybrid} & AO & Ethereum~\cite{wood2014ethereum}
        & Y & N & Nakamoto & \multirow{2}{*}{Y}\\
         &  & AO & Ronin chain~\cite{roninchain} & N & Y & BFT & \\
        \hline
        \multirow{2}{*}{Illuvium~\cite{illuvium}} & \multirow{2}{*}{hybrid} & AO & Ethereum~\cite{wood2014ethereum}
        & Y & N & Nakamoto & -\\
         &  & GP & AWS & - & -& - & -\\
        \hline
        \multirow{3}{*}{Stepn~\cite{stepn}} & \multirow{3}{*}{hybrid} & AO & Solana~\cite{solana}
        & Y & N & BFT & \multirow{3}{*}{Y}\\
         &  & AO & BSC
        & Y & Y(PoSA) & - & \\
         &  & AO & Ethereum
        & Y & N & Nakamoto & \\
        %  &  &  &  &  & & \\
        \hline
    \end{tabular}
    % \footnotetext[3]{a}
  \end{center}
  \end{minipage}
\end{table}
\subsection{Decentralized Data Storage}
\label{subsec:Decentralized data storage}
In this subsection, we discuss what to store and how to store it for improving the performance of decentralized data storage.

\noindent\textbf{Data in P2E Projects.}
Various types of data needed to be stored for P2E projects, \textit{e.g.}, game protocols \& designs of the universe, player actions, UGC, asset ownership, and various proofs. To explain these data in detail, we consider a Turing machine that stores these data and calculates the corresponding state of the game. This Turing machine is called the \textit{system} of the project. The system can contain centralized or decentralized subsystems storing different data. The game protocol and other designs of the game universe are sent to the system by the entrepreneur, to kick off the game universe. The game protocol defines how the players can play the game and what players can earn, \textit{e.g.}, how to battle, gain rewards, and breed Axie NFT in Axie Infinity~\cite{axie}. It also defines how to update the protocol itself. Other designs can be the appearance of the environment or avatars in the game, \textit{e.g.}, the character design of Axie character. 
Actions defined by the game protocol are sent by the players to play the game, \textit{e.g.}, registering the identity, purchasing assets, and controlling avatars. Then, the system can calculate asset ownership from the game protocol and the actions of players. UGC is content generated by the players. The content can be tokenized as NFT by the system, and become an asset that can be traded between players. For example, in The Sandbox~\cite{sandbox}, players can create items with its custom Voxel Engine on top of the Unity engine. In addition, the system can calculate succinct proofs for large data and store the proofs and the data differently to improve performance.
% Other than taking action in the universe, players can generate content.
% How these different data are stored in these subsystems will be discussed in Section 2.5.  
% The data sent to the system can be sent by various participants.
% state , data assets , environments, characters, updates  outside the consensus participants can send proofs to the system in order to prove the events that happen outside the system. For example,  

% some will affect the consensus of future states 
% For decentralized nodes, several techniques to improve performance
% proof

Next, we discuss how the above-mentioned data are stored in decentralized nodes. An intuitive way is to require each node to maintain a replica of all data. However, there will be a huge cost to store the growing data this way. Therefore, it is necessary to classify the data and store them differently. 
% \textcolor{red}{We will also introduce techniques to ensure data accessibility when the data are stored only in part of the nodes. }
According to whether it is required for nodes to reach a consensus, data in P2E projects can be basically divided into two categories, necessary data, and unnecessary data. For example, the ownership of assets is necessary because when trading an asset, nodes should verify if the seller owns it. Instead, the raw data of UGC is unnecessary because nodes can use its hash value to represent the data during the consensus. This will consume less storage and network resources. Also, some intermediate process data is unnecessary after they are aggregated. The necessary data is stored by all nodes participating in the consensus, thus easy to access during the consensus process. On the contrary, the data unnecessary for the consensus can only be saved by part of the nodes. The data inaccessibility issue caused by distributed storage will be discussed in Section~\ref{subsec:Attacks}. 

% We distinguish data by whether they are necessary for the nodes to reach a consensus to improve the performance of the implementation. For example, the own？ership of an asset is necessary for further consensus, because when trading the asset nodes should verify the ownership of the seller to execute the trading process otherwise the asset will be sold by someone who does not own the asset. On the other hand, the raw data of UGC are unnecessary once the system calculates the hash of the data because the nodes can use the hash to refer to the data in the consensus protocol. The hash of the data will consume less storage and network resources. Also, some intermediate processes are unnecessary after they are aggregated. The necessary data for the consensus are stored in all the nodes participating in the consensus protocol and thus they will be easy to access during the consensus process. The data unnecessary for the consensus can be stored in some of the nodes instead of all the nodes.
% other centralized or decentralized subsystems.

In addition, we discuss some techniques to enable secure decentralized storage in Appendix~\ref{sec:Secure Decentralized Data Storage}, including data structure, upload and download process, and incentives for nodes that provide storage. They are different from those of centralized storage because the decentralized nodes may be malicious, fail, or hard to find, for not having a registered address. 
% Replication of the data content addressing encryption incentive
% In the following paragraph

\subsection{Consensus}
\label{subsec:Consensus}
% As mentioned above, data can be stored in decentralized nodes. 
In order to have the same view of the game universe, nodes run consensus protocols to maintain consistency, during which they communicate with each other to reach continuous unanimous agreements on the states. Consensus protocols should satisfy safety and liveness~\cite{castro1999practical} so that participants can hold consistent states at any time and their inputs to the system will eventually act on the state.
% we require the states calculated by each node are the same. However, different nodes may store conflict states, because the network condition can be different between nodes and some nodes can even behave maliciously. To solve the problem, nodes run consensus protocols to maintain consistency. In consensus protocols, nodes communicate with each other to reach continuous unanimous agreements of the state. Consensus protocols satisfy safety and liveness~\cite{castro1999practical} so that participants can reach consistent states at any time and their input to the system will eventually act on the state.
The consensus of a P2E project can be categorized in terms of whether the project relies on a public blockchain, how the nodes join the network (\textit{i.e.}, permissioned or permissionless), and the protocol, \textit{i.e.}, how each node decides on the state.

\noindent\textbf{Public or Project-specific Chain.} 
% Whether the project relies on a public blockchain or a project-specific chain is used can affect the stability of the system.
P2E projects may use a public blockchain or a project-specific chain. Different ways can affect the stability of the system. For projects that run on a project-specific blockchain, the entrepreneur or DAO can control the parameters of the chain, such as transaction fees, the number of nodes, and the type of consensus. For example, Gala Games~\cite{gala} allows at most 50,000 consensus nodes and anyone who wants to join the consensus should buy a Founder’s Node License, the price of which is set by the entrepreneur. However, using a project-specific blockchain means that the consensus rewards are endogenous tokens of the chain. If this niche currency is not attractive to nodes, the security and decentralization of the consensus can not be ensured, making the system vulnerable. To bootstrap the project-specific chain, an unpredictable genesis block~\cite{garay2018bootstrapping} should also be prepared for the project. This block can either be provided by a trusted centralized party or a decentralized party, such as an existing blockchain. For example, the Ronin sidechain is boosted by a block of Ethereum. There are also plenty of P2E projects using public chains. For example, hundreds of P2E projects are based on Solana chain~\cite{SolanaNFTGames}. To maintain low transaction fees, high throughput, and fast transaction finalization, various methods are used, such as Layer-2 methods~\cite{gudgeon2020sok}, sharding~\cite{wang2019sok,yu2020survey}, and DAG-based blockchains~\cite{wang2020sok}. However, some of these methods may induce security problems, which will be discussed in Section~\ref{subsec:Attacks}. 

% However, it may be vulnerable, because miners are often rewarded by the project token. If there is not enough incentive for nodes, the security, and decentralization of the consensus can not be ensured.

\noindent\textbf{Permissioned or Permissionless.} In terms of how the nodes join the network, there can be permissionless and permissioned settings. For the latter, its identity management makes the confirmation speed faster and the realization easier, but meanwhile, the consensus nodes are more likely to collude with each other, posing a threat to the consensus security. On the other hand, consensus with permissionless nodes is considered to be more decentralized and transparent. P2E projects with permissioned designs usually adopt proof of authority (PoA) as the consensus algorithm while the consensus nodes are determined by the entrepreneur. For example, in the Ronin chain of Axie, the entrepreneur selects trusted nodes to build the consensus. Nodes can also be added upon the approval of the existing consensus nodes. In the permissionless setting, nodes use proof of work (PoW)~\cite{gervais2016security}, PoS, Delegated Proof-of-Stake (DPoS)~\cite{snider2018delegated} to elect voters and leaders~\cite{gilad2017algorand, pass2016hybrid} to prevent Sybil attack and claim incentives. For example, in Gala Games~\cite{gala} players can purchase a Founder’s Node License, which is similar to PoS, and download the Gala Node Software to run the node. Running a Founder’s Node is motivated by project tokens. Permissioned consensus can anchor blocks in permissionless consensus. For example, Chromia~\cite{chromia} plans to record hashes of its blocks in Bitcoin and Ethereum.

\noindent\textbf{Consensus Protocol.} There are various protocols P2E projects can use to reach a consensus, \textit{e.g.}, Nakamoto protocol, Byzantine Fault Tolerant (BFT) protocol, or RAFT. Different protocols resist different proportions of malicious nodes to ensure security. In Nakamoto protocol, a leader is elected at each round to include inputs, vote for previous states with the most votes, and calculate the latest state. The protocol can maintain safety and liveness with 50\% malicious nodes for a synchronous network. For a partially synchronous model, where the upper bound in the delay of messages is unknown to the honest parties, the number of malicious nodes it can resist will decrease according to the delay~\cite{garay2015bitcoin}.
In Byzantine protocol, leaders and a voting committee are elected at each round~\cite{gilad2017algorand}. The leaders include inputs and calculate their states. Voters will initially vote for some of their received states and amplify the votes that they receive more than a threshold~\cite{castro1999practical}. The protocol can resist 33\% malicious nodes in the voting committee. 
% In Snowflake protocol, no leader will be elected, and every node can include inputs. For those conflict inputs, each node continues sampling some nodes and query their preferred states~\cite{rocket2018snowflake}. the number of malicious nodes it can resist
RAFT~\cite{ongaro2014search} can ensure safety in a fail-stop model and can not resist any number of Byzantine malicious nodes.

\subsection{Bridges and Oracles}
\label{subsec:Bridges and oracles}
The system of the project can be made up of several subsystems. Each subsystem can be a P2P network of decentralized nodes or a centralized server. For example, the nodes that participate in the consensus protocol and the nodes that respond to the data accessing request can be two groups in different networks. The system can also contain several groups of nodes that run consensus severally. This subsection introduces bridges and oracles that connect different decentralized subsystems.
% For example, there is a bridge built between the public chain Solana and decentralized storage Arweave.

% When two systems need communication, it can done by themselves, especially for centralized systems.
% intermediary
% same/different nodes

% The project can store them by themselves, or use platforms. There are NFT platforms like NiftyGateway, Opensea(filecoin, XFS\footnote{https://docs.xfs.tech/}). Some blockchains inherently provide the service to store them. 
% Oracles can be classified by their data source and 
% centralized software and centralized data source  
% decentralized software 
\noindent\textbf{Oracles.} Oracles are used to link consensus to other information sources, \textit{e.g.}, discussion forums, random numbers, order of transactions, economic information such as the price of the dollar, and any other external events. In P2E projects, the update of the game protocol is often discussed in some online forums, and players can vote for the proposals proposed in these forums. Random numbers generated in the consensus may be predictable to some adversaries, therefore more unpredictable random numbers should be generated by the oracle. Also, some of the consensuses have the temporarily centralized problem that the leader can determine part of the order of transactions, \textit{i.e.}, miner extractable values\cite{daian2020flash}. Therefore, oracles can be used to determine the order of transactions~\cite{kelkar2020order}.

The data source of the oracle can be software source (\textit{e.g.}, online databases, websites, blockchain mem pools) or hardware source (\textit{e.g.}, sensors)~\cite{beniiche2020study}. An oracle that uses a single source is considered to be centralized for it is vulnerable to single-point failure, while decentralized oracles aggregate the information from multiple sources~\cite{mammadzada2020blockchain}. There can be a hierarchical structure for a decentralized oracle, namely, a project can aggregate the information from different oracles, and an oracle can aggregate the information from several nodes that contract with different data sources to prevent malicious behaviors at any step. Nodes in the oracle can also form a consensus to validate the data~\cite{breidenbach2021chainlink}.

\noindent\textbf{Bridges.} Bridges are used to exchange or migrate assets between different consensuses~\cite{zamyatin2021sok}. There are many cross-chain P2E projects. For example, players can play Stepn in Solana, BNBchain, and Ethereum, which are called different realms in Stepn~\cite{StepnRealm}. Bridges across these chains help the participants react to the possible risks from the chains, such as transaction fee fluctuations, enabling them to choose their preferred chain at any time. In addition, for those project-specific chains, bridges are needed for users to exchange other commonly used cryptocurrencies for the project tokens. For example, the Ronin bridge is used to connect the Ronin chain and Ethereum so that players can buy Axie NFTs and other Axie tokens by ETH.

%The exchange process is that two players can exchange their assets in different chains with each other. 
Furthermore, Bridges enables the migration process, in which assets in one subsystem can be locked in the bridge accounts and some `wrapped' assets are issued in another subsystem, and if the `wrapped' assets are paid back, the original assets will be unlocked. So through this, players can migrate their assets from one chain to another. Projects can add rules on the usage of the `wrapped' assets. For example, in Stepn, Energy assets can only be shared across realms under specific conditions through Energy Bridge.

Bridges can be realized by a centralized entity. For example, Axie Ronin bridge is controlled by the entrepreneur of Axie project. Bridges can also be realized by a decentralized third party, such as chain relay or synchronization methods like HTLC~\cite{zamyatin2021sok}.
% Similar to the consensus, bridges can be consensus program-specific 
% program-specific bridge, trusted centralized third party bridge, decentralized bridge

\subsection{Hybrid Architecture}
\label{subsec:Hybrid architecture}
In the above sections, we mainly discussed the realization when all the data are stored in decentralized nodes, while many existing P2E projects use a hybrid architecture. In a hybrid architecture, there will be centralized servers storing and processing part of the data, and communicating with the decentralized subsystems of the project. A centralized server contains data storage and processing nodes owned by a single entity, \textit{e.g.}, entrepreneur, a third-party service provider like Amazon Web Services (AWS), and InterPlanetary file system (IPFS) pinning service providers such as Pinata~\cite{pinata} and Infura~\cite{infura}. There are several reasons to adopt a hybrid architecture. One reason is that the realization and the modification of the game protocol are easier for the entrepreneur when starting the project. Although the right to modify the protocol may harm players' interests, it is hard to design a perfect protocol that does not need subsequent modification, for there can be many uncertainties in the start-up period. Another reason is that the hybrid architecture can provide better performance because its requirements for the network throughput and data processing speed are much less. The main concern of the hybrid architecture is the failure of the centralized server. There can be inconsistency between the centralized server and the decentralized nodes due to attacks from an attacker or misbehavior of the entity who owns or rents the server, which will be discussed in Section~\ref{sec:Challenges}.
% , as in Figure~\ref{fig:hybrid_architecture}
% \begin{figure}
%     \centering
%     \includegraphics[width=0.6\linewidth]{figures/hybrid_architecture.png}
%     \caption{hybrid architecture}
%     \label{fig:hybrid_architecture}
% \end{figure}

Here, we give some examples of hybrid architecture that reflect different design principles. Usually, the most critical data like asset ownership, UGC, and the game protocol are stored in a decentralized way. In P2E projects, earning assets from the game is one of the most important features. Usually, the asset ownership will be stored by the decentralized nodes. For example, Illuvium~\cite{illuvium} is a collection game where players can control the avatars, explore the game universe and capture the encountered Illuvials. The ownership of Illuvials is written in an Ethereum smart contract. On the other hand, the project uses centralized AWS as its backend to store other data, such as the movement of the avatars. During the game, the action data of avatars are sent to the centralized server, while the trading data related to asset ownership are widely sent to the decentralized nodes. After that, the centralized server and decentralized nodes synchronize the latest state of the system. 
%\textcolor{red}{Players can send the actions of their avatars to the centralized server which stores the game protocol, and the centralized server can update the asset ownership to the decentralized nodes, while players can trade their assets by sending requests directly to the decentralized nodes, and the centralized server should catch up with the latest state in the decentralized nodes.} 
The design principle of this hybrid architecture is the transparency of the asset transfer, for the transfer process is realized in decentralized nodes. Therefore, players can monitor the centralized server's behavior and partially transfer their rights. For example, players can make necessary their authorization when the centralized server intends to move assets out of their accounts.

The accessibility and censorship resistance of UGC can be important in some projects. Therefore, the project can store UGC in decentralized nodes. For example, in The Sandbox, after players send their created content to the centralized server of the project, the server will calculate the hash of the content, tokenize the content, and send the raw data to decentralized storage nodes. The centralized server in The Sandbox is AWS and the decentralized storage is IPFS. Therefore, if the decentralized storage is properly designed, players can access the contents without the centralized server. The user front-end can also be used to tokenize the content and send the content to the decentralized nodes to ensure censorship resistance.
% For example, Decentraland is a game where players can claim ownership of virtual land \footnote{https://decentraland.org/whitepaper.pdf}. The land ownership is written in a smart contract in Ethereum.

The immutability of the game protocol is essential to a P2E project. For example, My Neighbor Alice~\cite{myneighboralice} emphasizes that for a game fully controlled by players, the logic of items should be stored in a decentralized way instead of only storing the ownership of the items. 

Some technologies can be used to improve the performance of the hybrid architecture. Trusted Execution Environments, \textit{e.g.}, Intel SGX~\cite{costan2016intel}, can be used to replace the centralized server. Then the system will be secure under the trust in the hardware and the institution that authenticates attestation keys~\cite{cheng2019ekiden}. User front-ends can be utilized to store and process data, as they are motivated to do so without incentives. For example, in Decentraland, players can play games on the servers hosted by landowners. 

\section{Challenges}
\label{sec:Challenges}
In this section, we discuss security \& societal challenges of the projects. In Section~\ref{subsec:Attacks}, we discuss security challenges. In Section~\ref{subsec:Economic Challenges}, we discuss economic challenges. In Section~\ref{subsec:Governance challenges}, we discuss governance challenges.
\subsection{Security Challenges}
\label{subsec:Attacks}
% \begin{itemize}
%     \item rug pull
%     \item require() rollback
%     \item front run
%     \item loopholes in smart contract
%     \item bridge
%     \item chain
%     \begin{itemize}
%         \item private chain (Axie
%         \item layer2 papers
%         \item PoS
%     \end{itemize}
%     % \item oracle (random, )
% \end{itemize}
\textbf{ICO Rug Pull.}
During the ICO phase, investors take the risk that the entrepreneur may abruptly shut down the project after raising money. Many rug pulls have been observed in various blockchain projects~\cite{gan2021initial}. In the field of P2E games, there are also such scams. Along with the shutdown of the project, the entrepreneur deletes their social accounts and transfers their raised cryptocurrency to money launderers. Towards this problem, one possible approach for the investors is to judge the quality of a project by the sold fraction of the project token~\cite{chod2021theory}. If the proportion of tokens issued in the ICO stage is quite high, such a project is likely to have the risk of rug pull.

\noindent\textbf{Private Key Compromise.}
To obtain a large amount of money, the common accounts in the game universe, bridges of multi-chain projects, and cryptocurrency wallets are mostly targeted, especially when they are not fully decentralized. Attackers may try to get hold of the participants' private keys to steal tokens in their wallets. Actually, private keys can be compromised by bribing the key owner, hacking or phishing, or just because of the reuse of private keys. 

In addition, they may attempt to compromise the private keys of consensus nodes to control the entire chain directly. To pursue efficiency, many P2E projects use private chains or layer-2 scaling schemes instead of the public chain for recording most of the game operations. In a private chain, several specified nodes take responsibility for the consensus. The private chain is tied to other chains through two-way bridges for cross-chain coordination. If larger than 50\% nodes are corrupted in a private chain, the attacker can take control of the whole chain. As a result, they can corrupt all the accounts, including the game treasury and all bridges, and take the tokens therein. For example, Axie Infinity uses the Ronin, which is a private sidechain linked to Ethereum. On March 29, 2022, Axie Infinity reported a loss of over \$625 million caused by an attack on the Ronin sidechain. The attacker hacked to get the private keys of over half of Ronin nodes and take the cryptocurrency to Ethereum through a bridge.

Likewise, Layer-2 scaling schemes also face this kind of problem. Layer-2 schemes can provide P2E games with high throughput by allowing offline processes. Layer-2 protocols rely on a set of computational nodes. A classic solution is that any computational node could provide periodic checkpoints to the main public chain while others can challenge it by offering fraud proofs. For example, the Polygon blockchain uses this kind of solution and several P2E projects (\textit{e.g.}, Arc8 by GAMEE, Crazy Defense Heroes, and Pegaxy) are built on it. In such schemes, if all computational nodes are corrupted, the attacker can also operate the system arbitrarily~\cite{kalodner2018arbitrum,khalil2018commit}.

%Private chains are those with their own consensus and are tied to other chains by two-way bridges. For example, Axie Infinity uses the Ronin, which is a private sidechain linked to Ethereum. Layer2 scaling schemes are those that provide periodic checkpoints to a public chain. For example P2E projects on the polygon, such as Arc8 by GAMEE, Crazy Defense Heroes, and Pegaxy. If larger than 50\% nodes are corrupted in a private chain or all the nodes are corrupted in a layer2 scheme like Arbitrum~\cite{kalodner2018arbitrum} or Nocust~\cite{khalil2018commit}, the attacker can take control of all the accounts, including game treasury and bridges to the private chain. On March 29, 2022, Axie Infinity reported a loss of over \$625 million caused by an attack on the Ronin sidechain. The attacker hacked to get the key of over half of Ronin nodes and take the crypto-currency to Ethereum through a bridge.
\noindent\textbf{Smart Contract Vulnerabilities.}
As a blockchain project, the P2E game deploys smart contracts to deal with the game rules like token flows and the minting mechanism. Researchers have found that there are multiple security issues in the publicly deployed smart contracts, which may be maliciously exploited by the attackers~\cite{atzei2017survey,perez2021smart}. Although extensive literature investigates vulnerability detection and security analysis~\cite{schneidewind2020ethor}, some blockchain projects still get attacked due to the improper design or implementation of smart contracts, and P2E games are no exception. 

In P2E projects, contracts transferring NFTs are usually the target. For example, in the initial CryptoPunks contract, the ETH paid to buy the punk NFT was wrongly sent to the buyer instead of the NFT seller. As a result, attackers take advantage of the loophole to earn NFTs for free. There are also other designs of the smart contract that may cause an unexpected loss of other honest users in the community.
\begin{itemize}
    \item \textbf{Rollback of randomness.} In Ethereum virtual machine, some functions (\textit{e.g.}, \texttt{require} and \texttt{revert} function) can be invoked to rollback the transaction. Therefore when randomness is involved to determine the result of a request, the sender can roll back the transaction if the result is not satisfactory. For example, in Cryptozoon~\cite{cryptozoon}, the player can buy an egg to hatch a ZOAN NFT, which will be randomly given a Rarity Level. When initiating such a transaction, players can decide to roll back it if the obtained level is not desirable. The solution to this kind of attack can be checking whether buying function is called by a contract, and prohibiting the contract call.
    
\item \textbf{Front running.} The transparent nature of blockchain makes DApps highly vulnerable to the front-running attack~\cite{eskandari2019sok}. For example, players in PolkaMonster sell their NFTs by setting a price in the marketplace. It is observed that the seller successfully earned a lot of money by front running. The attacker first set a low price for his NFT to attract a buyer. After that, he could monitor transactions in the network. When detecting that someone issued a transaction to buy this NFT,\footnote{See \url{https://bscscan.com/tx/0x1a31bfc4d3c4a726c931cb784f9e79606d62996b6251c9ad959e5b2e6621fd9e} for this buying transaction, the gas fee of which is 5 Gwei.} he immediately updated the price and set a higher gas fee for this updating transaction.\footnote{See \url{https://bscscan.com/tx/0xb9ec7e204f186660a377beeee8e9223107f46d8718314fbfdc4133580311442e} for this updating transaction, the gas fee of which is 7.7 Gwei.} As a result, the price updating was executed first and the buyer finally paid a much higher price. The solution to this kind of attack can be letting the buyer set an acceptable price range or encrypting the transactions.
\end{itemize}

% https://www.tuoniaox.com/news/p-512604.html
% https://cryptozoon.io/CryptoZoon-WhitePaper.pdf
    
% 基于EOS的底层逻辑，WAX采用了DPoS（Delegated Proof of Stake）机制。在 WAX 上，21 个活跃的 WAX 公会和 36 个备用的 WAX 公会每 0.5 秒生成一个区块。理论上，公链每秒可以处理3000笔交易。实际上，WAX 在其高度时每秒完成 1,418 笔交易，链上 gas 费用几乎为零。
% Games on WAX: https://playtoearn.net/blockchaingames/WAX/All-Genre/All-Status/All-Device/All-NFT/All-PlayToEarn/All-FreeToPlay
\noindent\textbf{NFT Content Inaccessibility.}
In most of the existing P2E projects, the NFT content is stored in centralized storage, or decentralized storage with improper incentives. For example, they are stored in AWS, centralized Pinning Service providers, or IFPS without incentive mechanisms, as mentioned in Section~\ref{sec:Implementation}. In this case, the content may be lost~\cite{NFTsMissing}. Although we can store the content locally, in P2E projects, there are various data that need to ensure accessibility, such as game environments and character designs generated by the entrepreneur. For example, the costume of a player's avatar should be accessed by other players to realize its social function. There will also be content generated by players that need access when players are playing the game. For example, one of the use cases of the land in Decentraland is advertising. There is no meaning if the advertisement can not be accessed publicly.

% or malicious behaviors when the game universe is just created

\subsection{Economic Challenges}
\label{subsec:Economic Challenges}
\textbf{Token Design.}
Token designs affect the economy. As mentioned in Section~\ref{subsec:Tokens}, some projects use a single token for multiple purposes, while others use different tokens for corresponding purposes. If fewer tokens are used, the token design will be simpler. But on the other hand, the use of multiple tokens allows the system to respond flexibly to various conditions and participants with different preferences. 
% However, to fully attract participants with different preferences in different situations of the projects, corresponding tokens should be designed with proper inflation or deflation. 
For example, the incentive-for-player tokens and funding tokens can be different in terms of stability and issuance schedule. To attract myopic players with the network effect, the price of incentive-for-player tokens should be stable. Otherwise, the decline of the token price may cause collective abandonment of the game. To maintain a steady user growth rate, the incentive tokens can be designed to be more valuable when the system suffers slow growth. Instead, funding tokens can be unstable to reward those visionary investors. 
%\textcolor{red}{Therefore, there can be incentive tokens more valuable at a lower player growth rate to maintain a stable growth rate, while funding tokens can be unstable but more valuable when there are more players to reward the investors for their visionary investment.} 
In terms of the issuance method, there can be either a scheduled supply or an unscheduled supply. Scheduled supply means the total amount of the token is fixed or the issuance of every period is scheduled. Unscheduled supply is that the issuance amount will depend on the activity of players in the game universe, \textit{e.g.}, tokens can be issued every time a player wins a mini-game and in this way, the more winning events happen, the larger total supply of tokens there will be. Tokens with scheduled supply are often used for funding, while tokens with unscheduled supply will be more attractive to those players with a less available cost for playing.

\noindent\textbf{Irregular Issuance.}
With more and more players playing the game, more tokens will be earned by players, especially those unscheduled utility tokens that can be easily earned by playing but cannot be used for staking or governance. As a result, the price of tokens may decrease. To maintain stability, designers consider reducing those tokens' supply or increasing consumption. For example, the utility token of Axie Infinity (\textit{i.e.}, SLP) experienced a rapider and earlier price drop than AXS, so Axie Infinity reduces the amount of SLP players can earn in its season 20 update. In addition, to increase the demand of these tokens, new events in the universe can be developed. For example, Axie Infinity designs NFT Runes \& Charms which can be crafted by consuming SLP in its new Origin season. 

% To attract new players, tokens can be designed of players with different risk preference. For example, in Thetan Arena, players can either start for free and earn less, or first pay and earn more.

%  The price of token\_A and token\_X will also decrease $\delta P^a_t$ and $\delta P^x_t$. otherwise, supply will be larger than demand. In ICO, the price of token\_A will be usually lower than the price after the universe is constructed, for the investor bear the risk. If the investor is improvident, he may sell out his token. Also we can not expect all these investors will hold the ICO tokens for a long time.

Therefore, tokenomics should be carefully designed to balance those impacts. There can be some swap mechanisms to make scheduled and unscheduled tokens able to be exchanged in the universe. Besides, it can enable these two kinds of tokens to have similar usage so that players can strategically choose which token to use, bringing a dynamic adjustment.

% Therefore, the tokenomics should be carefully designed to balance those impacts or let players strategically choose the optimal portfolio by designing some swap mechanism. For example, making scheduled tokens and unscheduled tokens able to be exchanged in the universe or some scheduled tokens and unscheduled tokens may have similar usage so that players can strategically choose which token to use.

\noindent\textbf{Friction When Attracting New Players.} High gas fees for the blockchains and the unfamiliarity of potential players with blockchain technologies can cause friction when the project wants to attract new players. To solve the problem, the project can implement the game simultaneously in public chains, private chains, and centralized servers, and build bridges between them so that players can start the game with low fees.

\noindent\textbf{Wash Trading.}
Users of the blockchains can easily lend flash loan~\cite{wang2020towards} and create multiple accounts. Therefore, anyone can create multiple accounts and transfer tokens among them, which is called wash trading. For those fungible tokens, auto market makers (AMM) can prevent them from wash trading, because the trading price cannot be set directly by the seller. However, there is no effective mechanism to prevent NFT wash trading nowadays~\cite{WashTrading}. 

Wash trading has a significant influence on P2E projects, for it leads to larger trading volume and price distortion. Some statics websites, such as CoinMarkCap, rank P2E projects by the transaction volume, and the higher the volume is, the more visible the project will be to the potential investors and players~\cite{cong2021crypto}. In addition, the price of a set of similar NFTs is calculated by the average of their sale prices. Therefore, wash trading at a high price can affect the evaluated value of these NFTs. Wash trading also enables money laundry. For example, the ICO process of a malicious project may be utilized to become part of a money laundry process~\cite{MoneyLaundry}.

\subsection{Governance Challenges}
\label{subsec:Governance challenges}
The P2E projects will go through a centralization phase, a decentralization phase, and an intermediate transition period. In the decentralization phase, a proposal-and-majority-voting scheme is widely used, namely, any participant in the game universe can put forward proposals and vote for them, and their vote is weighted by their tokens like Apecoin~\cite{apecoin}. We discuss the challenges in the decentralization phase in terms of voting security, voting privacy, governance efficiency, governance fairness, and the tragedy of the commons.
% \begin{itemize}
%         \item voting security \& privacy: voter \& miner, especially for those have program-specific chain
%         \item governance efficiency
%         \item fairness
%         \item Tragedy of the Commons~\cite{faysse2005coping}
%     \end{itemize}
%     \item before decentralization, there are also challenges
%     \begin{itemize}
%         \item ICO rug pull
%         \item tokens that the entrepreneur do not possess: e.g. updating game versions
%     \end{itemize}

\noindent\textbf{Governance Security in the Centralization Phase.}
% \textbf{entrepreneur}
% In some of the P2E projects, there is no assurance when players can vote for the governing issues, although that brand themselves as decentralized ecosystems. Also, to realize decentralization, it should be ensured that the weight of governance token that can be obtained by playing or public sale is larger than that entrepreneur own and obtain by staking.
In the traditional joint equity system, everyone holds the same equity. Therefore, it is reasonable for a centralized entity that possesses the majority of the equity to conduct the governance. However, in P2E projects, there are tokens that are only possessed by players, \textit{e.g.}, SLP in Axie Infinity. Since the governor (namely, the entrepreneur) does not hold such tokens, he may make decisions that harm the interests of those token holders. For example, the entrepreneur may update the game universe, limit the usage of these tokens, and issue new tokens to attract new players but devalue the original tokens.

In addition, most P2E projects did not identify when and how to enter the decentralization phase in the smart contract, for it is hard to predict the best time to do so. Therefore, the entrepreneur can arbitrarily postpone the decentralization phase. 

% project tokens that the entrepreneur do not posses
\noindent\textbf{Governance Security in the Transition Period.}
During the transition period from the centralization phase to the ideal decentralization process, the voting weight of the entrepreneur will decrease while that of players is the reverse. Their interests may conflict during the shift of duty. For example, before the voting weight of the entrepreneur is down to less than half, he can make most of the decisions. Therefore, the interests of the players can not be ensured.
%  The weight are represented by the tokens.The weight are represented by the tokens.
% \textbf{proposal \& voting security}

\noindent\textbf{Proposal \& Voting Security.}
After entering the decentralization stage, the consensus nodes ensure proposal and voting security, which should behave as neutral entities to collect proposals, tally the votes, and manage the encryption and delegation of votes~\cite{zhang2018treasury}. Here, encryption is for voting privacy, and delegation is for governance efficiency, which will be discussed in the rest of this section. However, the consensus nodes may refuse to include the votes or proposals they are opposed to, especially for those projects that own their specific consensus. For example, in some projects, only miners can vote for the proposals. In this case, the rights and interests of other players that do not run consensus nodes can not be guaranteed.

% In addition, proposals are usually proposed outside the consensus.
\noindent\textbf{Voting Privacy.}
What an individual voter voted on should not be revealed to anyone during the voting process or even after the process. Privacy during the voting process is to mitigate the malicious behavior when tallying the votes, which may be realized by the commit-and-reveal scheme~\cite{hardwick2018voting}. The privacy after the voting process is to prevent bribery and collusion, making the voting protocol coercion-resistance~\cite{delaune2009verifying}. To realize the privacy decryption, trustees~\cite{cortier2019belenios} or voting committee~\cite{zhang2018treasury} should be elected for the decryption of the aggregated votes. However, the privacy of voting is not realized prevalently in the cryptocurrency community, \textit{e.g.}, voting of Bitcoin improvement proposal~\cite{bips}, and Apecoin DAO Governance~\cite{apecoin}. Since the pseudonymity of the blockchain does not ensure privacy, the voting scheme should be designed carefully.
% There two kinds of privacy of voting incl privacy within the process of voting, i.e. during the process of voting, anyone cannot know who vote for what but after, and i.e.
% process privacy and ultimate privacy~\cite{zhang2018treasury}

\noindent\textbf{Governance Inefficiency.}
The governance inefficiency comes from two facts players can join or leave the community freely, and not all the voters are always online.

When players can join or leave the community freely, voters are dynamically changing. Therefore, different or even conflict policies may be implemented. 

For the voters' offline problem, enough voting time should be set to make sure token holders notice the vote has begun and fully discuss the proposals. Otherwise, governance may be taken over by an attacker, a centralized entrepreneur, or other unexpected token holders.
On February 12, 2022, Build Finance DAO encountered a Governance Takeover Attack. The attacker succeeded in the takeover by having a large enough vote in favor of the proposal to take control of the Build token contract~\cite{GovernanceTakeover}. Although the risk can be mitigated by setting a voting threshold, \textit{i.e.}, how many votes are needed to approve a proposal, or restricting the amount of token transfer of a single proposal. However, these solutions will take time and harm the efficiency of the governance. Delegating the votes or liquid democracy~\cite{blum2016liquid,zhang2018treasury} can also be a solution to the inefficiency. In liquid democracy, the delegation can be terminated at any time, having little harm to decentralization from a long-term perspective.
% To reduce the risk of governance being taken over by an attacker, the following methods can be adopted:\\
% 1. Voting threshold on  needs to be properly set.\\
% 2. Voting periods need to be long enough \\
% 3. Restrictions on proposals and behavior like minting tokens and draining the treasury can be set.\\
% 4. 
% 5. The transfer of the governance token can be tracked and the voting weight can be reduced after the token is traded. For example, non-transferable NFT, soulbond~\cite{weyl2022decentralized} can be used in the P2E project so that only the player who earns the token by playing can vote by the token.
% flash loan

\noindent\textbf{Governance Fairness.}
% \textbf{unipolarity}
% In the governance of P2E projects, the traditional fairness that every account has the same voting weight can not be satisfied, because of Sybil attacks that account can be created without any costs. 
The governance of P2E projects does not adopt the method that each account has the same voting weight because accounts can be easily created without any cost. Usually, votes are weighted by project tokens to realize the fairness that the voting weight is proportional to the time, effort, or money a player spent on the project. However, for a token allocation proposal concerning how to allocate the tokens in the treasury, players with a large number of tokens may vote for the decision that benefits them. Thus, the proposal that distributes more tokens to those possessing more voting tokens will be more likely to be accepted. In this way, rich people will become richer, which is opposite to the sense of fairness. To mitigate the unfairness, we can restrict the distribution of the governance token. For example, governance tokens can only be obtained by playing the game and further soulbound~\cite{weyl2022decentralized} to the account that obtains them, making it difficult to be traded.

In addition, P2E projects usually have multiple governance tokens. When voting for a proposal, weights are set to each token, and a voter's voting weight is the weighted addition of the amount of each token he possesses. The weight is set either exactly the same as or different from the real-time exchange rate. If the weight is different from the real-time exchange rate, malicious voters may exchange for a cheaper token with larger weights. On the other hand, if the weight is approximately the same as the real-time exchange rate, players possessing the token with a small market cap will be in an inferior position. For example, if there are two groups using different tokens in the game universe, all decisions will be made by the group with tokens of a larger market cap under the proposal-and-majority-voting scheme. To meet better fairness with multiple governance tokens, other voting schemes can be used. For example, when deciding on a division of some resources, instead of the majority voting for a proposal of a particular distribution, calculating a distribution based on the preference aggregation of voters will be fairer~\cite{lackner2020perpetual,fain2016core}. 

% Therefore, if the governance weight of these tokens are different, the  of token holders with less weight, especially those holders of non governance token, may be hurt. For example, the proposal of increasing the staking  of governance tokens are more likely to be approved. To mitigate these, subDAO can be used to vote for different issues with different weights of tokens.
% flash loan

% https://cryptobriefing.com/build-finance-dao-suffers-governance-takeover-attack/
%account bound
\noindent\textbf{Tragedy of the Commons.}
When governance is conducted by the token holders, the long-term benefit of the project can not be ensured, because some of the token holders can be myopic. For example, they can refuse to increase the budget on the incentives for new players that may have long-term merit on the value of the tokens. To mitigate this kind of effect, we can: 1) fix the budget ratio of different categories~\cite{zhang2018treasury}, including marketing, development, and other necessary components for a project to become sustainable; 2) force the voters to deposit their stake for a long time to make sure the voters hold the token for a long enough period before or after voting so that the voters will consider the long-term return and the financial risk of the attacker will be increased, \textit{e.g.}, in stepN, users will get higher voting power by locking GMT~\cite{stepnGovernance}; 3) use subDAOs to vote for different issues with proper weights of tokens. For example, in Yieldguild, issues of different games will be voted for with different subDAO tokens.

\subsection{Ownership of Digital Assets}
One of the features of P2E projects is that players can have true ownership of the assets in the game universe. However, the ownership is not well defined. Since the entrepreneur that creates the game universe is the most considered central entity. Whether the universe can be accessed and run without the entrepreneur or whether the entity that can modify the game is chosen by the community of players may be one of the rules to test if the project is decentralized. In many current projects, the hybrid consensus is used, and the codes of the game universe are not open-sourced. If the entrepreneur becomes malicious and denial of service, the tokens players hold in the decentralized consensus will not have the same value as before. However, anyone can pirate the game if the source code is revealed. Therefore, the decentralization process can be seen as the entrepreneur selling the game universe to the decentralized players via the smart contract.

% If the source code is reveal, the type of games may be restricted

% TEE
\section{Conclusion}
In this paper, we discuss the implementation, token models, and governance in the play-to-earn projects and propose challenges and some solutions in terms of attacks on the implementation layer, the economic failure that may happen in player incentive, and the difficulty of governance in the play-to-earn projects.

\newpage
\bibliographystyle{splncs04}
\bibliography{reference}

\begin{thebibliography}{100}
\providecommand{\url}[1]{\texttt{#1}}
\providecommand{\urlprefix}{URL }
\providecommand{\doi}[1]{https://doi.org/#1}

\bibitem{aavegotchi}
Aavegotchi litepaper, \url{https://wiki.aavegotchi.com/} Accessed October 6,
  2022

\bibitem{alienworlds}
Alien worlds - blockchain technical blueprint,
  \url{https://docs.google.com/document/d/1JiA97Y3JZMcC6HG2VPXEiZDd7UtA5yJSRUY2DQ5VSRI/edit#heading=h.k1xajz1gmpx4}
  Accessed October 6, 2022

\bibitem{apecoin}
Apecoin dao, \url{https://apecoin.com/governance#voting-mechanism} Accessed
  October 6, 2022

\bibitem{arweave}
Arweave whitepaper, \url{https://www.arweave.org/files/arweave-whitepaper.pdf}
  Accessed October 6, 2022

\bibitem{AxiePeak}
Axie infinity: Pernicious pyramid scheme or gaming breakthrough?,
  \url{https://www.forbes.com/sites/leeorshimron/2022/08/13/axie-infinity-pernicious-pyramid-scheme-or-gaming-breakthrough/?sh=1b25657074b3}
  Accessed October 16, 2022

\bibitem{axie}
Axie infinity whitepaper, \url{https://whitepaper.axieinfinity.com} Accessed
  October 6, 2022

\bibitem{battleworld}
Battle world white paper, \url{https://whitepaper.battleworld.game/} Accessed
  October 6, 2022

\bibitem{MoneyLaundry}
The biggest threat to trust in cryptocurrency: Rug pulls put 2021
  cryptocurrency scam revenue close to all-time highs,
  \url{https://blog.chainalysis.com/reports/2021-crypto-scam-revenues/}
  Accessed October 16, 2022

\bibitem{bsc}
Binance smart chain white paper,
  \url{https://github.com/bnb-chain/whitepaper/blob/master/WHITEPAPER.md}
  Accessed October 6, 2022

\bibitem{bips}
Bitcoin improvement proposals, \url{https://github.com/bitcoin/bips/} Accessed
  October 16, 2022

\bibitem{GovernanceTakeover}
The build finance dao has been the target of a hostile governance takeover,
  \url{https://twitter.com/finance_build/status/1493223190071554049?s=20&t=ZXnf1ZAAGFpqQhDUBbbmtw}
  Accessed October 16, 2022

\bibitem{chromia}
Chromia,
  \url{https://chromia.com/documents/Chromia-_-Platform-white-paper2019.pdf}
  Accessed October 6, 2022

\bibitem{counterparty}
Counterparty homepage, \url{https://counterparty.io/} Accessed October 6, 2022

\bibitem{cryptozoon}
Cryptozoon whitepaper, \url{https://cryptozoon.io/CryptoZoon-WhitePaper.pdf}
  Accessed October 6, 2022

\bibitem{decentraland}
Decentraland whitepaper, \url{https://decentraland.org/whitepaper.pdf} Accessed
  October 6, 2022

\bibitem{filecoin}
Filecoin whitepaper, \url{https://filecoin.io/filecoin.pdf} Accessed October 6,
  2022

\bibitem{gala}
Gala games, \url{https://support.gala.games/} Accessed October 6, 2022

\bibitem{galanode}
Gala games node ecosystem roadmap,
  \url{https://gogalagames.medium.com/node-pre-proposal-node-rewards-for-network-contribution-ea3c29cdce46}
  \&
  \url{https://gogalagames.medium.com/the-gala-games-node-ecosystem-9760d8156af7}
  Accessed October 6, 2022

\bibitem{CryptoKittiesPeak}
Gamefi explained: How to earn money through blockchain gaming,
  \url{https://www.binance.com/en-NG/blog/all/gamefi-explained-how-to-earn-money-through-blockchain-gaming-421499824684903406}
  Accessed October 16, 2022

\bibitem{stepnGovernance}
Governance \& ecosystem fund,
  \url{https://whitepaper.stepn.com/governance-and-ecosystem-fund} Accessed
  October 16, 2022

\bibitem{hive}
Hive white paper, \url{https://hive.io/whitepaper.pdf} Accessed October 6, 2022

\bibitem{illuvium}
Illuvium whitepaper, \url{https://docs.illuvium.io/whitepaper/technology}
  Accessed October 6, 2022

\bibitem{immutable}
Immutable x white paper,
  \url{https://assets.website-files.com/62535c6262b90afd768b9b26/6304335ed396fd9c8d8dfe5e_Immutable\%20X\%20Whitepaper.pdf}
  Accessed October 6, 2022

\bibitem{dappradar}
Industry overview, \url{https://dappradar.com/industry-overview} Accessed
  October 16, 2022

\bibitem{infura}
Infura document, \url{https://docs.infura.io/infura} Accessed October 6, 2022

\bibitem{klaytn}
Klaytn documentation, \url{https://docs.klaytn.foundation/} Accessed October 6,
  2022

\bibitem{makerdao}
Makerdao whitepaper, \url{https://makerdao.com/en/whitepaper} Accessed October
  6, 2022

\bibitem{minesofdalarnia}
Mines of dalarnia lightpaper,
  \url{https://www.minesofdalarnia.com/assets/MoD-Litepaper-updated-27-Oct.pdf}
  Accessed October 6, 2022

\bibitem{mobox}
Mobox document, \url{https://faqen.mobox.io/} Accessed October 6, 2022

\bibitem{myneighboralice}
My neighbor alice whitepaper,
  \url{https://static1.squarespace.com/static/601a9e78af724f7a3932fd5f/t/6197c7a3208b111b3333c707/1637337019912/My+Neighbor+Alice+Whitepaper+-+final.pdf}
  Accessed October 6, 2022

\bibitem{pinata}
Pinata website, \url{https://www.pinata.cloud/} Accessed October 6, 2022

\bibitem{polygon}
Polygon lightpaper, \url{https://polygon.technology/lightpaper-polygon.pdf}
  Accessed October 6, 2022

\bibitem{radiocaca}
Radiocaca, \url{https://www.radiocaca.com/} Accessed October 6, 2022

\bibitem{readon}
Readon whitepaper, \url{https://whitepaper.readon.me/what-is-readon/overview}
  Accessed October 6, 2022

\bibitem{roninchain}
Ronin network documentation, \url{https://docs.roninchain.com/docs/intro}
  Accessed October 6, 2022

\bibitem{sandbox}
The sandbox whitepaper,
  \url{https://installers.sandbox.game/The_Sandbox\_Whitepaper\_2020.pdf}
  Accessed October 6, 2022

\bibitem{solana}
Solana documentation, \url{https://docs.solana.com/cluster/synchronization}
  Accessed October 6, 2022

\bibitem{spellsofgenesis}
Spells of genesis homepage, \url{https://spellsofgenesis.com/} Accessed October
  6, 2022

\bibitem{splinterlands}
Splinterlands documentation, \url{https://docs.splinterlands.com/} Accessed
  October 6, 2022

\bibitem{stepn}
Stepn whitepaper, \url{https://whitepaper.stepn.com/} Accessed October 6, 2022

\bibitem{StepnRealm}
Stepn's realms, \url{https://whitepaper.stepn.com/realm} Accessed October 16,
  2022

\bibitem{SolanaNFTGames}
Top solana nft games by market cap 2022,
  \url{https://chainplay.gg/chain/solana/} Accessed October 16, 2022

\bibitem{ufogaming}
Ufogaming, \url{https://ufogaming.io/} Accessed October 6, 2022

\bibitem{wax}
Wax protocol white paper,
  \url{https://github.com/worldwide-asset-exchange/whitepaper} Accessed October
  6, 2022

\bibitem{wemix}
Wemix whitepaper, \url{https://che.wemixnetwork.net/hm/rsc/whitepaper.pdf}
  Accessed October 6, 2022

\bibitem{AxieStorm}
What is axie infinity? the play-to-earn nft game taking crypto by storm,
  \url{https://decrypt.co/resources/what-is-axie-infinity-the-play-to-earn-nft-game-taking-crypto-by-storm}
  Accessed October 16, 2022

\bibitem{NFTsMissing}
Yes, your nfts can go missing—here's what you can do about it,
  \url{https://decrypt.co/62037/missing-or-stolen-nfts-how-to-protect} Accessed
  October 16, 2022

\bibitem{ateniese2020proof}
Ateniese, G., Chen, L., Etemad, M., Tang, Q.: Proof of storage-time:
  Efficiently checking continuous data availability. Cryptology ePrint Archive
  (2020)

\bibitem{atzei2017survey}
Atzei, N., Bartoletti, M., Cimoli, T.: A survey of attacks on ethereum smart
  contracts (sok). In: International conference on principles of security and
  trust. pp. 164--186. Springer (2017)

\bibitem{benet2014ipfs}
Benet, J.: Ipfs-content addressed, versioned, p2p file system. arXiv preprint
  arXiv:1407.3561  (2014)

\bibitem{benet2017proof}
Benet, J., Dalrymple, D., Greco, N.: Proof of replication. Protocol Labs, July
  \textbf{27}, ~20 (2017)

\bibitem{beniiche2020study}
Beniiche, A.: A study of blockchain oracles. arXiv preprint arXiv:2004.07140
  (2020)

\bibitem{blum2016liquid}
Blum, C., Zuber, C.I.: Liquid democracy: Potentials, problems, and
  perspectives. Journal of Political Philosophy  \textbf{24}(2),  162--182
  (2016)

\bibitem{breidenbach2021chainlink}
Breidenbach, L., Cachin, C., Chan, B., Coventry, A., Ellis, S., Juels, A.,
  Koushanfar, F., Miller, A., Magauran, B., Moroz, D., et~al.: Chainlink 2.0:
  Next steps in the evolution of decentralized oracle networks. Chainlink Labs
  (2021)

\bibitem{canetti2019fiat}
Canetti, R., Chen, Y., Holmgren, J., Lombardi, A., Rothblum, G.N., Rothblum,
  R.D., Wichs, D.: Fiat-shamir: from practice to theory. In: Proceedings of the
  51st Annual ACM SIGACT Symposium on Theory of Computing. pp. 1082--1090
  (2019)

\bibitem{castro1999practical}
Castro, M., Liskov, B., et~al.: Practical byzantine fault tolerance. In: OsDI.
  vol.~99, pp. 173--186 (1999)

\bibitem{chen2017improved}
Chen, Y., Li, H., Li, K., Zhang, J.: An improved p2p file system scheme based
  on ipfs and blockchain. In: 2017 IEEE International Conference on Big Data
  (Big Data). pp. 2652--2657. IEEE (2017)

\bibitem{cheng2019ekiden}
Cheng, R., Zhang, F., Kos, J., He, W., Hynes, N., Johnson, N., Juels, A.,
  Miller, A., Song, D.: Ekiden: A platform for confidentiality-preserving,
  trustworthy, and performant smart contracts. In: 2019 IEEE European Symposium
  on Security and Privacy (EuroS\&P). pp. 185--200. IEEE (2019)

\bibitem{chod2021theory}
Chod, J., Lyandres, E.: A theory of icos: Diversification, agency, and
  information asymmetry. Management Science  \textbf{67}(10),  5969--5989
  (2021)

\bibitem{cong2021crypto}
Cong, L.W., Li, X., Tang, K., Yang, Y.: Crypto wash trading. arXiv preprint
  arXiv:2108.10984  (2021)

\bibitem{cortier2019belenios}
Cortier, V., Gaudry, P., Glondu, S.: Belenios: a simple private and verifiable
  electronic voting system. In: Foundations of Security, Protocols, and
  Equational Reasoning, pp. 214--238. Springer (2019)

\bibitem{costan2016intel}
Costan, V., Devadas, S.: Intel sgx explained. Cryptology ePrint Archive  (2016)

\bibitem{daian2020flash}
Daian, P., Goldfeder, S., Kell, T., Li, Y., Zhao, X., Bentov, I., Breidenbach,
  L., Juels, A.: Flash boys 2.0: Frontrunning in decentralized exchanges, miner
  extractable value, and consensus instability. In: 2020 IEEE Symposium on
  Security and Privacy (SP). pp. 910--927. IEEE (2020)

\bibitem{delaune2009verifying}
Delaune, S., Kremer, S., Ryan, M.: Verifying privacy-type properties of
  electronic voting protocols. Journal of Computer Security  \textbf{17}(4),
  435--487 (2009)

\bibitem{eskandari2019sok}
Eskandari, S., Moosavi, S., Clark, J.: Sok: Transparent dishonesty:
  front-running attacks on blockchain. In: International Conference on
  Financial Cryptography and Data Security. pp. 170--189. Springer (2019)

\bibitem{fain2016core}
Fain, B., Goel, A., Munagala, K.: The core of the participatory budgeting
  problem. In: International Conference on Web and Internet Economics. pp.
  384--399. Springer (2016)

\bibitem{fisch2019tight}
Fisch, B.: Tight proofs of space and replication. In: Annual International
  Conference on the Theory and Applications of Cryptographic Techniques. pp.
  324--348. Springer (2019)

\bibitem{gan2021initial}
Gan, J., Tsoukalas, G., Netessine, S.: Initial coin offerings, speculation, and
  asset tokenization. Management Science  \textbf{67}(2),  914--931 (2021)

\bibitem{garay2015bitcoin}
Garay, J., Kiayias, A., Leonardos, N.: The bitcoin backbone protocol: Analysis
  and applications. In: Annual international conference on the theory and
  applications of cryptographic techniques. pp. 281--310. Springer (2015)

\bibitem{garay2018bootstrapping}
Garay, J.A., Kiayias, A., Leonardos, N., Panagiotakos, G.: Bootstrapping the
  blockchain, with applications to consensus and fast pki setup. In: IACR
  International Workshop on Public Key Cryptography. pp. 465--495. Springer
  (2018)

\bibitem{gervais2016security}
Gervais, A., Karame, G.O., W{\"u}st, K., Glykantzis, V., Ritzdorf, H., Capkun,
  S.: On the security and performance of proof of work blockchains. In:
  Proceedings of the 2016 ACM SIGSAC conference on computer and communications
  security. pp. 3--16 (2016)

\bibitem{gilad2017algorand}
Gilad, Y., Hemo, R., Micali, S., Vlachos, G., Zeldovich, N.: Algorand: Scaling
  byzantine agreements for cryptocurrencies. In: Proceedings of the 26th
  symposium on operating systems principles. pp. 51--68 (2017)

\bibitem{gudgeon2020sok}
Gudgeon, L., Moreno-Sanchez, P., Roos, S., McCorry, P., Gervais, A.: Sok:
  Layer-two blockchain protocols. In: International Conference on Financial
  Cryptography and Data Security. pp. 201--226. Springer (2020)

\bibitem{hardwick2018voting}
Hardwick, F.S., Gioulis, A., Akram, R.N., Markantonakis, K.: E-voting with
  blockchain: An e-voting protocol with decentralisation and voter privacy. In:
  2018 IEEE International Conference on Internet of Things (iThings) and IEEE
  Green Computing and Communications (GreenCom) and IEEE Cyber, Physical and
  Social Computing (CPSCom) and IEEE Smart Data (SmartData). pp. 1561--1567.
  IEEE (2018)

\bibitem{kalodner2018arbitrum}
Kalodner, H., Goldfeder, S., Chen, X., Weinberg, S.M., Felten, E.W.: Arbitrum:
  Scalable, private smart contracts. In: 27th USENIX Security Symposium (USENIX
  Security 18). pp. 1353--1370 (2018)

\bibitem{kelkar2020order}
Kelkar, M., Zhang, F., Goldfeder, S., Juels, A.: Order-fairness for byzantine
  consensus. In: Annual International Cryptology Conference. pp. 451--480.
  Springer (2020)

\bibitem{khalil2018commit}
Khalil, R., Zamyatin, A., Felley, G., Moreno-Sanchez, P., Gervais, A.:
  Commit-chains: Secure, scalable off-chain payments. Cryptology ePrint Archive
   (2018)

\bibitem{kiayias2017ouroboros}
Kiayias, A., Russell, A., David, B., Oliynykov, R.: Ouroboros: A provably
  secure proof-of-stake blockchain protocol. In: Annual international
  cryptology conference. pp. 357--388. Springer (2017)

\bibitem{lackner2020perpetual}
Lackner, M.: Perpetual voting: Fairness in long-term decision making. In:
  Proceedings of the AAAI conference on artificial intelligence. vol.~34, pp.
  2103--2110 (2020)

\bibitem{mammadzada2020blockchain}
Mammadzada, K., Iqbal, M., Milani, F., Garc{'\i}a-Ba{~n}uelos, L.,
  Matulevi{\v{c}}ius, R.: Blockchain oracles: A framework for blockchain-based
  applications. In: International Conference on Business Process Management.
  pp. 19--34. Springer (2020)

\bibitem{moin2020sok}
Moin, A., Sekniqi, K., Sirer, E.G.: Sok: A classification framework for
  stablecoin designs. In: International Conference on Financial Cryptography
  and Data Security. pp. 174--197. Springer (2020)

\bibitem{oliveira2018token}
Oliveira, L., Zavolokina, L., Bauer, I., Schwabe, G.: To token or not to token:
  Tools for understanding blockchain tokens  (2018)

\bibitem{ongaro2014search}
Ongaro, D., Ousterhout, J.: In search of an understandable consensus algorithm.
  In: 2014 USENIX Annual Technical Conference (Usenix ATC 14). pp. 305--319
  (2014)

\bibitem{pagnia1999impossibility}
Pagnia, H., G{\"a}rtner, F.C., et~al.: On the impossibility of fair exchange
  without a trusted third party. Tech. rep., Citeseer (1999)

\bibitem{pass2016hybrid}
Pass, R., Shi, E.: Hybrid consensus: Efficient consensus in the permissionless
  model. Cryptology ePrint Archive  (2016)

\bibitem{perez2021smart}
Perez, D., Livshits, B.: Smart contract vulnerabilities: Vulnerable does not
  imply exploited. In: 30th USENIX Security Symposium (USENIX Security 21). pp.
  1325--1341 (2021)

\bibitem{saleh2021blockchain}
Saleh, F.: Blockchain without waste: Proof-of-stake. The Review of financial
  studies  \textbf{34}(3),  1156--1190 (2021)

\bibitem{schneidewind2020ethor}
Schneidewind, C., Grishchenko, I., Scherer, M., Maffei, M.: ethor: Practical
  and provably sound static analysis of ethereum smart contracts. In:
  Proceedings of the 2020 ACM SIGSAC Conference on Computer and Communications
  Security. pp. 621--640 (2020)

\bibitem{snider2018delegated}
Snider, M., Samani, K., Jain, T.: Delegated proof of stake: features \&
  tradeoffs. Multicoin Cap  \textbf{19} (2018)

\bibitem{steichen2018blockchain}
Steichen, M., Fiz, B., Norvill, R., Shbair, W., State, R.: Blockchain-based,
  decentralized access control for ipfs. In: 2018 Ieee international conference
  on internet of things (iThings) and ieee green computing and communications
  (GreenCom) and ieee cyber, physical and social computing (CPSCom) and ieee
  smart data (SmartData). pp. 1499--1506. IEEE (2018)

\bibitem{vakilinia2022incentive}
Vakilinia, I., Wang, W., Xin, J.: An incentive-compatible mechanism for
  decentralized storage network. arXiv preprint arXiv:2208.09937  (2022)

\bibitem{vidal2022new}
Vidal-Tom{\'a}s, D.: The new crypto niche: Nfts, play-to-earn, and metaverse
  tokens. Finance Research Letters p. 102742 (2022)

\bibitem{WashTrading}
von Wachter, V., Jensen, J.R., Regner, F., Ross, O.: {NFT} wash trading:
  Quantifying suspicious behaviour in {NFT} markets. CoRR
  \textbf{abs/2202.03866} (2022), \url{https://arxiv.org/abs/2202.03866}

\bibitem{wang2020towards}
Wang, D., Wu, S., Lin, Z., Wu, L., Yuan, X., Zhou, Y., Wang, H., Ren, K.:
  Towards understanding flash loan and its applications in defi ecosystem.
  arXiv preprint arXiv:2010.12252  (2020)

\bibitem{wang2019sok}
Wang, G., Shi, Z.J., Nixon, M., Han, S.: Sok: Sharding on blockchain. In:
  Proceedings of the 1st ACM Conference on Advances in Financial Technologies.
  pp. 41--61 (2019)

\bibitem{wang2020sok}
Wang, Q., Yu, J., Chen, S., Xiang, Y.: Sok: Diving into dag-based blockchain
  systems. arXiv preprint arXiv:2012.06128  (2020)

\bibitem{werner2021sok}
Werner, S.M., Perez, D., Gudgeon, L., Klages-Mundt, A., Harz, D., Knottenbelt,
  W.J.: Sok: Decentralized finance (defi). arXiv preprint arXiv:2101.08778
  (2021)

\bibitem{weyl2022decentralized}
Weyl, E.G., Ohlhaver, P., Buterin, V.: Decentralized society: Finding web3's
  soul. Available at SSRN 4105763  (2022)

\bibitem{wilkinson2014storj}
Wilkinson, S., Boshevski, T., Brandoff, J., Buterin, V.: Storj a peer-to-peer
  cloud storage network  (2014)

\bibitem{wood2014ethereum}
Wood, G., et~al.: Ethereum: A secure decentralised generalised transaction
  ledger. Ethereum project yellow paper  \textbf{151}(2014),  1--32 (2014)

\bibitem{yu2020survey}
Yu, G., Wang, X., Yu, K., Ni, W., Zhang, J.A., Liu, R.P.: Survey: Sharding in
  blockchains. IEEE Access  \textbf{8},  14155--14181 (2020)

\bibitem{zamyatin2021sok}
Zamyatin, A., Al-Bassam, M., Zindros, D., Kokoris-Kogias, E., Moreno-Sanchez,
  P., Kiayias, A., Knottenbelt, W.J.: Sok: Communication across distributed
  ledgers. In: International Conference on Financial Cryptography and Data
  Security. pp. 3--36. Springer (2021)

\bibitem{zhang2018treasury}
Zhang, B., Oliynykov, R., Balogun, H.: A treasury system for cryptocurrencies:
  Enabling better collaborative intelligence. Cryptology ePrint Archive  (2018)

\end{thebibliography}
\appendix
\newpage
\section{Secure Decentralized Data Storage}
\label{sec:Secure Decentralized Data Storage}
In this section, we discuss techniques to store data securely on decentralized nodes in terms of data structure, upload and download process, and incentives for nodes that provide storage.

\noindent\textbf{Data Structure.}
To make sure to get the correct data, data usually break into blocks and form Merkle DAGs by hashes of each block, \textit{e.g.}, content addressing in InterPlanetary file system (IPFS) ~\cite{benet2014ipfs}. Therefore, the data will be immutable, and all the replications of the same data will have the same address. 
To make the searching process easier, tags or metadata on data format can be appended to the hash.
Data can also be encoded as erasure code~\cite{chen2017improved} to make the replication of data more efficient. Privacy and access management~\cite{steichen2018blockchain} can be ensured by end-to-end encryption as in Storj~\cite{wilkinson2014storj}. Access management for NFT content may be essential because, in P2E projects, royalties may need to be paid to NFT creators. 

\noindent\textbf{Upload and Download Process.}
To upload and download the data, storage-providing nodes can be found by a market or dissemination between nodes. For example, in Filecoin~\cite{filecoin}, there is a market matching the data owner and storage-providing nodes. The data owner can set the price and choose how many replications are needed for the data. On the other hand, in Arweave~\cite{arweave}, nodes disseminate the data and choose whether to store the data when a data owner uploads the data. During the download process, nodes also disseminate the download request to neighboring nodes to find the nodes that possess the data.
% their ranking system is used to behave these nodes.

\noindent\textbf{Incentives.}
The incentive for nodes to store the data should be set properly. Otherwise, unpopular files may be trashed, and nodes may deny the service to the client~\cite{vakilinia2022incentive}. The data owner can pay fees either when requesting storage or requesting accessing the data. When paying for accessing the data, an atomic exchange is needed between the data and the fee. However, it is impossible without a trusted third party~\cite{pagnia1999impossibility}. Erasure code~\cite{wilkinson2014storj} and payment channels may be used to mitigate the risk.

To implement proper incentives, proofs are needed. According to different needs, there can be two kinds of proof, proof of replication/storage of the data or access to certain data. Proof of replication (PoRep)~\cite{benet2017proof} is dedicating unique physical storage, which requires the prover to provide processed data, and the process takes time to prevent the prover from fetching the corresponding data from another storage provider. 
On the other hand, proof of access (PoA) lets the storage provider decide whether to store the data.
Therefore, those projects that attach more importance to data accessibility can choose PoRep, while PoA may be more economic.
Since using whole data for proof and continuous periodical checking bring too much burden to the network. Several methods can be used to generate succinct proof, \textit{e.g.}, probabilistic sampling using Merkle tree, non-interactive scheme~\cite{ateniese2020proof}~\cite{canetti2019fiat}, zk-snark or other encodings~\cite{fisch2019tight}.
These proofs can either check by the users or contracts on the consensus ledger.
% Non-interactive proof using methods similar to the Fiat-Shamir method.
% These proofs can workload proof as a side-product

\section{Tables}
\begin{sidewaystable}[htbp]
  \caption{The purposes of tokens of the P2E projects. The Require ownership means players should own the tokens to unlock some features in the game. These tokens can also be seen as paying before playing in rewarded after playing. Exchanging tokens at a fixed exchange rate is considered a playing process in this paper. Therefore, the tokens that can be gained by a fixed rate exchange are classified in the Incentive-for-players column.}\label{tab:token}
\scriptsize
  \begin{center}
    \begin{tabular}{|c|c|c|c|c|c|c|c|c|c|}
        \hline
        \multirow{2}{*}{projects} & \multirow{2}{*}{Funding} & Incentive for & \multirow{2}{*}{Payment} & Require & \multirow{2}{*}{Voting} & Incentive for & Tokenize & Incentive for & Stake proof\\
         & & players &  & ownership* &  & investors & UGC & implementation & for consensus\\
        \hline
        My Neighbor Alice~\cite{myneighboralice} & ALICE & ALICE & ALICE & lands\&items & ALICE & ALICE & items & - & -\\
        %  &  &  & anchor blocks in PoW chains &  & \\
        \hline
        \multirow{2}{*}{Wemix Play~\cite{wemix}} & \multirow{2}{*}{WEMIX} & WEMIX, & WEMIX, & \multirow{2}{*}{-} & \multirow{2}{*}{-} & \multirow{2}{*}{WEMIX} & \multirow{2}{*}{-} & \multirow{2}{*}{WEMIX} & \multirow{2}{*}{-}\\
         &  & GameToken & GameToken &  &  & &&&\\
         \hline
        \multirow{2}{*}{Splintershards~\cite{splinterlands}} & SPS, card, & \multirow{2}{*}{SPS, DEC} & SPS, DEC, & \multirow{2}{*}{card} & \multirow{2}{*}{SPS} & \multirow{2}{*}{SPS} & \multirow{2}{*}{-} & \multirow{2}{*}{SPS} & \multirow{2}{*}{-}\\
         & CREDIT &  & CREDIT &  &  & &&&\\
         \hline
        \multirow{2}{*}{Alien Worlds~\cite{alienworlds}} & \multirow{2}{*}{Trilium} & Trilium, & \multirow{2}{*}{Trilium} & \multirow{2}{*}{Core NFTs} & \multirow{2}{*}{Trilium} & \multirow{2}{*}{Trilium} & Landowner & \multirow{2}{*}{-} & \multirow{2}{*}{-}\\
         &  & NFTs\footnote{including Core NFTs and Landowner-offered NFTs} &  &  &  &  & -offered NFTs &&\\
        \hline
        \multirow{2}{*}{Gala Games~\cite{gala}} & \multirow{2}{*}{GALA} & \multirow{2}{*}{GameToken} & GALA & \multirow{2}{*}{-} & Founder’s& \multirow{2}{*}{-} & \multirow{2}{*}{-} & \multirow{2}{*}{GALA} & Founder’s \\
         & & & GameToken &  & Node license &  & & & Node license \\
        %   &  & & &  &  & (entrepreneur)&\\
        \hline
        \multirow{2}{*}{The Sandbox~\cite{sandbox}} & \multirow{2}{*}{SAND} & SAND & SAND, GEMS & \multirow{2}{*}{LAND} & \multirow{2}{*}{SAND} & SAND, GEMS & \multirow{2}{*}{ASSETS} & - & - \\
         &  & ASSETS & CATALYSTS &  &  & CATALYSTS &  &  & \\
        \hline
        \multirow{2}{*}{Decentraland~\cite{decentraland}} & \multirow{2}{*}{MANA} & \multirow{2}{*}{MANA} & \multirow{2}{*}{MANA} & \multirow{2}{*}{LAND} & MANA, LAND & \multirow{2}{*}{-} & \multirow{2}{*}{UGC} & \multirow{2}{*}{-} & \multirow{2}{*}{-} \\
         &  &  &  &  & NAMES &  &  &  & \\
         \hline
        \multirow{2}{*}{Axie Infinity~\cite{axie}} & \multirow{2}{*}{AXS} & AXS, SLP & \multirow{2}{*}{AXS, SLP} & \multirow{2}{*}{Axie NFT} & \multirow{2}{*}{AXS} & \multirow{2}{*}{AXS} & \multirow{2}{*}{-} & \multirow{2}{*}{-} & \multirow{2}{*}{-} \\
         &  & Axie NFT &  &  &  &  &  &  & \\
        \hline
        Illuvium~\cite{illuvium} & ILV & Illuvium NFT & ETH~\cite{wood2014ethereum} & Illuvium NFT & ILV & Promo NFT & - & - & - \\
        \hline
        \multirow{2}{*}{Stepn~\cite{stepn}} & \multirow{2}{*}{GMT} & GMT, GST & \multirow{2}{*}{GMT, GST} & \multirow{2}{*}{Sneaker NFT} & \multirow{2}{*}{GMT} & \multirow{2}{*}{-} & \multirow{2}{*}{-} & \multirow{2}{*}{-} & \multirow{2}{*}{-} \\
         &  & Sneaker NFT &  &  &  &  &  &  & \\
        \hline
        RadioCaca~\cite{radiocaca} & RACA & NFTs\footnote{including Metamon Egg, portion, and diamond} & RACA, NFTs & Metamon NFT & RACA & NFTs & - & - & - \\
        \hline
        \multirow{2}{*}{UFO Gaming~\cite{ufogaming}} & \multirow{2}{*}{UFO} & \multirow{2}{*}{Genesis NFT} & Plasma & \multirow{2}{*}{Genesis NFT} & \multirow{2}{*}{-} & UFO, & \multirow{2}{*}{-} & \multirow{2}{*}{-} & \multirow{2}{*}{-}\\
         &  &  & Points &  &  & Plasma Points &  &  & \\
        \hline
        \multirow{2}{*}{Aavegotchi~\cite{aavegotchi}} & Aavegotchi & \multirow{2}{*}{GHST, NFTs\footnote{including Portals and Wearables}} & \multirow{2}{*}{GHST, NFTs} & \multirow{2}{*}{Aavegotchi NFT} & \multirow{2}{*}{GHST} & \multirow{2}{*}{-} & \multirow{2}{*}{-} & \multirow{2}{*}{-} & \multirow{2}{*}{-}\\
         & NFT ,GHST &  &  &  &  &  &  &  & \\
        \hline
        % PlayMining &  &  &  &  &  & \\
        % \hline
        \multirow{2}{*}{MOBOX~\cite{mobox}} & \multirow{2}{*}{MBOX, GEM} & \multirow{2}{*}{MBOX, MEC} & MBOX, & \multirow{2}{*}{MOMO NFT} & \multirow{2}{*}{MBOX} & \multirow{2}{*}{MBOX}
        & \multirow{2}{*}{-} & \multirow{2}{*}{-} & \multirow{2}{*}{-}\\
         &  &  & GEM, MEC &  &  &  &  &  & \\
        \hline
        \multirow{2}{*}{Mines of Dalarnia~\cite{minesofdalarnia}} & \multirow{2}{*}{DAR, Land} & DAR, & DAR, & \multirow{2}{*}{Land} & \multirow{2}{*}{DAR} & \multirow{2}{*}{-} & \multirow{2}{*}{-} & \multirow{2}{*}{-} & \multirow{2}{*}{-}\\
         &  & mineral & mineral &  &  &  &  &  & \\
        \hline
        Battle world~\cite{battleworld} & BWO & BWO & BWO & NFTs & - & BWO & - & - & -\\
        \hline
        % \hline
        % \hline
        % \hline
    \end{tabular}
  \end{center}
\end{sidewaystable}

\begin{sidewaystable}[htbp]
\caption{The implementations of the P2E projects introduced in their whitepaper or documents.}
    \label{tab:implementation}
\scriptsize
  \begin{center}
    \begin{tabular}{|c|c|c|c|c|c|c|c|}
        \hline
        \multirow{2}{*}{projects} & \multirow{2}{*}{architecture} & \multirow{2}{*}{data} & \multirow{2}{*}{implementation} & public/& permissioned/ & consensus  & \multirow{2}{*}{bridge}\\
         &  & & &program-specific& permissionless & protocol &\\
        \hline
        My Neighbor Alice~\cite{myneighboralice} & decentralized & all\footnote{including asset ownership, UGC, game protocol and its execution} & Chromia~\cite{chromia} & public & permissioned & BFT & -\\
        %  &  &  & anchor blocks in PoW chains &  & \\
        \hline
        \multirow{2}{*}{Wemix Play~\cite{wemix}} & \multirow{2}{*}{decentralized} & asset\footnote{ including Wemix token} ownership & Klaytn~\cite{klaytn} & public & permissioned & BFT & \multirow{2}{*}{Wemix bridge}\\
         &  & others\footnote{including Game token~\cite{wemix} ownership, game protocols and their execution} & Wemix chain & specific & permissioned & RAFT & \\
         \hline
        \multirow{2}{*}{Splintershards~\cite{splinterlands}} & \multirow{2}{*}{decentralized} & asset ownership & WAX~\cite{wax} & public & permissionless(DPoS) & BFT & Splinterlands\\
         &  & game protocol\footnote{including game protocols and their execution} & Hive~\cite{hive}& public & permissionless(DPoS) & - & WAX bridge\\
         \hline
        \multirow{3}{*}{Alien Worlds~\cite{alienworlds}}& \multirow{3}{*}{decentralized} &asset ownership  & BSC~\cite{bsc} & public & permissioned(PoSA) & - & Cross-chain\\
        &  & main game protocol & WAX~\cite{wax} & public & permissionless(DPoS) & BFT & reconciling\\
        &  & NFT images & IPFS & - & - & - & -\\
        \hline
        \multirow{2}{*}{Gala Games~\cite{gala}} & \multirow{2}{*}{hybrid} & asset ownership & Ethereum~\cite{wood2014ethereum} & public & permissionless & Nakamoto & \multirow{2}{*}{-}\\
         & & other game content & Gala Node Ecosystem~\cite{galanode} & specific & permissionless & monitored by Gala Game& \\
        %   &  & & &  &  & (entrepreneur)&\\
        \hline
        \multirow{3}{*}{The Sandbox~\cite{sandbox}} & \multirow{3}{*}{hybrid} & asset ownership & Ethereum~\cite{wood2014ethereum}
        & public & permissionless & Nakamoto & -\\
         &  & UGC & IPFS~\cite{benet2014ipfs} & - & - & -&-\\
          &  & game protocol & AWS\footnote{Amazon Web Services} & - & - & -&-\\
        \hline
        \multirow{3}{*}{Decentraland~\cite{decentraland}} & \multirow{3}{*}{hybrid} & asset ownership & Ethereum~\cite{wood2014ethereum}
        & public & permissionless & Nakamoto & -\\
         &  & UGC & BitTorrent/IPFS & - & -& - & -\\
         &  & game protocol & User front-end & - & -& - & -\\
         \hline
        \multirow{2}{*}{Axie Infinity~\cite{axie}} & \multirow{2}{*}{hybrid} & asset ownership & Ethereum~\cite{wood2014ethereum}
        & public & permissionless & Nakamoto & \multirow{2}{*}{Ronin bridge}\\
         &  & asset ownership & Ronin chain~\cite{roninchain} & specific & permissioned & BFT & \\
        \hline
        \multirow{2}{*}{Illuvium~\cite{illuvium}} & \multirow{2}{*}{hybrid} & asset ownership & Ethereum~\cite{wood2014ethereum}
        & public & permissionless & Nakamoto & -\\
         &  & game protocol & AWS & - & -& - & -\\
        \hline
        \multirow{3}{*}{Stepn~\cite{stepn}} & \multirow{3}{*}{hybrid} & asset ownership & Solana~\cite{solana}
        & public & permissionless & BFT & \multirow{3}{*}{Energy Bridge}\\
         &  & asset ownership & BSC
        & public & permissioned(PoSA) & - & \\
         &  & asset ownership & Ethereum
        & public & permissionless & Nakamoto & \\
        %  &  &  &  &  & & \\
        \hline
        \multirow{2}{*}{RadioCaca~\cite{radiocaca}}& hybrid & asset ownership & BSC & public & permissioned(PoSA) & - & \multirow{2}{*}{-}\\
         & hybrid & asset ownership & Ethereum & public & permissionless & Nakamoto & \\
        \hline
        UFO Gaming~\cite{ufogaming} & hybrid & asset ownership & ETH(Immutable X~\cite{immutable}) & public & permissionless & Nakamoto &- \\
        \hline
        Aavegotchi~\cite{aavegotchi} & hybrid & asset ownership & ETH(Polygon~\cite{polygon}) & public & permissionless & Nakamoto & -\\
        \hline
        % PlayMining &  &  &  &  &  & \\
        % \hline
        \multirow{3}{*}{MOBOX~\cite{mobox}} & \multirow{3}{*}{hybrid} & asset ownership & BSC & public & permissioned(PoSA) & - & \multirow{3}{*}{-}\\
         &  & UGC & IFPS & - & - & - & \\
          &  & - & MOBOX Chain \footnote{under development(https://faqen.mobox.io/ecosystem/mobox-chain)} & - & - & - & \\
        \hline
        Mines of Dalarnia~\cite{minesofdalarnia} & hybrid & asset ownership & BSC & public & permissioned(PoSA) & - & - \\
        \hline
        Battle world~\cite{battleworld} & hybrid & asset ownership & ETH(Polygon) & public & permissionless & Nakamoto &- \\
        \hline
        % \hline
        % \hline
        % \hline
    \end{tabular}
    % \footnotetext{a}
  \end{center}
\end{sidewaystable}
\newpage




% LNCS Homepage, \url{http://www.springer.com/lncs}. Last accessed 4
% \end{thebibliography}
\end{document}